\documentclass[12pt,pre,aps,epsf,floats,onecolumn]{revtex4}

\usepackage{amssymb}
\usepackage{graphicx}

\begin{document}

\markboth{Stefano Battiston, Joao F. Rodrigues, Hamza Zeytinoglu}{The Network of
Inter-Regional Direct Investment Stocks Across Europe}

\title{The Network of Inter-Regional Direct Investment Stocks across Europe}

\author{STEFANO BATTISTON}

\author{JOAO F. RODRIGUES}

\affiliation{Chair for Systems Design,
ETH Zurich\\
D-MTEC, Kreuzplatz 5, CH-8092 Zurich, Switzerland\\
sbattiston@ethz.ch; jrodrigues@ethz.ch}

\author{HAMZA ZEYTINOGLU}

\affiliation{Clarifix Ltd., Suite 302, 95 Wilton Road, London SW1V 1BZ, UK\\
hamza@clarifix.net}

\begin{abstract}
We propose a methodological framework to study the dynamics of inter-regional
investment flow in Europe from a Complex Networks perspective, an approach with
recent proven success in many fields including economics. In this work we study
the network of investment stocks in Europe at two different levels: first, we
compute the inward-outward investment stocks at the level of firms, based on
ownership shares and number of employees; then we estimate the inward-outward
investment stock at the level of regions in Europe, by aggregating the ownership
network of firms, based on their headquarter location. Despite the intuitive value
of this approach for EU policy making in economic development, to our knowledge
there are no similar works in the literature yet. In this paper we focus on
statistical distributions and scaling laws of activity, investment stock and
connectivity degree both at the level of firms and at the level of regions. In
particular we find that investment stock of firms is power law distributed with an
exponent very close to the one found for firm activity. On the other hand
investment stock and activity of regions turn out to be log-normal distributed. At
both levels we find scaling laws relating investment to activity and connectivity.
In particular, we find that investment stock scales with connectivity in a similar
way as has been previously found for stock market data, calling for further
investigations on a possible general scaling law holding true in economical
networks.
\end{abstract}


\maketitle

\section{Introduction}\label{sec:Intro}

In this work we study the network of investment stocks in Europe at two different
levels of graining, a finer and a coarser level. We start by studying the network of
investment stocks between individual firms and we then proceed to study the network of
investment stocks between European regions.
At each level we focus on two specific aspects of such networks: on one
hand the statistical distributions of activity, investment stock and connectivity
degree. On the other, the scaling relations between such quantities.
In this respect this work is related to previous ones in Complex Networks, in economic
networks, and also in scaling laws in industrial economics.

Several authors within the Complex Networks scientific community have started to focus
today on the geographical aspects of empirical networks like for instance the
world wide air traffic network \cite{Barrat/Barth/Vesp:05:SpatConstrInAirNet,Guimera/Mossa/Turtschi/Amaral:05:AirNet}.

Closer to economics, some authors have studied the so called World Trade Web
(WTW), i.e. the network of import-export trade among countries in the world
\cite{Garlaschelli/Loffredo:04:WTW,Serrano/Boguna:03:WTW}. Also, statistical
properties of ownership networks have been studied so far only in a few works, and
in any case with no focus on geographical aspects. To name a few of the works in
this area, a study on the firm ownership network of Germany found it to exhibit
small world properties \cite{Kogut and Walker 1999}. In the US stock markets,
power law distribution of connectivity degree and scaling relation between degree
and invested volume have been found \cite{Garlaschelli/Battiston/al:05:SFTopo}.
However, it has been observed that network structures may differ from each other
in terms of control concentration and still look similar from the point of view of
the degree distribution or the investment volume distribution
\cite{Battiston:04:InnerStructCapitNets}.

In the economic literature, some works consider geographical embedding of economic
networks. It has been argued that domestic rivalry and geographic industry
concentration are especially important in creating dynamic clusters
\cite{Jakobsen/Onsager:04:HeadOfficeLocation}. From a more general perspective, some
authors assume the existence of a global
world-economy in which, since its inception in the sixteenth century, the
periphery is assigned the function of supplying the core with cheap labor and raw
materials, while the core has the role of producing manufactured goods, which
require capital intensive high technology \cite{Wallerstein:74:ModernWorldSystemI}.

Statistical distributions in firm demography are well documented: firm size
\cite{Axtell:01:Science,Sutton:97}, firm growth
\cite{Fujiwara/DiGuilmi/Aoyama/Gallegati:03:ParetoZipf}, firm debts
\cite{DelliGatti/DiGuilmi/Gallegati:03} have been studied in several countries and
contexts. However, in general, interactions are poorly considered in empirical
studies. Firms are not isolated, but rather depend on the interaction with other
ones through supply, ownership, partnership relations etc. Such relations should
then play a role in the statistical properties observed at a macroscopic scale.
Also, there exists an interaction between firms and banks through loans, and it
has been found that the distribution of the so called "bad debts" (debts of firms
towards the bank when they go bankrupt) is power-law distributed. Another type of
interaction which is not well documented are the investment stocks of firms in
other firms. In this paper we find a power law distribution for investment stock
with an exponent very close to the one found for firm activity (measured in number
of employees, see section \ref{sec:Methods}). Such a finding suggests a scaling
law between investment and activity that we also investigate and discuss. It is
relevant to mention here the relationship between empirically observed macroscopic
distributions and possible microscopic processes that may be responsible for them.
In literature about the emerging of scaling laws in firm demography, power law
distributions are usually obtained as a result of a random multiplicative process
affecting the size of each firm. Now, it is known from the works of Kesten in the
70'ies that under some general conditions, a combination of random multiplicative
and additive process can give rise to power law distributions
\cite{Kesten:73:RandomDiff}. Even a random multiplicative process with a lower
reflecting barrier (representing for example a bankruptcy threshold below which a
firm disappears and a new one is created) gives rise to a power law distribution.
Instead, a pure random multiplicative process gives rise to a log-normal
distribution \cite{Sornette:98}. However, in all these processes, interaction
between firms is not considered. Among the few works addressing the issue, we
mention one in which a firm bankruptcy affects indirectly other firms through the
interest rate of the central bank \cite{Delli Gatti/al:03:NewApprBusinessFluct}.
In this paper we start filling such gap by focusing on the distribution and the
scaling properties of investments of firms in other firms. Our aim is to
contribute to the understanding of how interdependency between firms gives rise to
well defined statistical distributions both at the level of firms as at the level
of regions.

\subsection{Foreign Direct Investments and Inter-Regional Direct Investments}
Concerning investment there is a tendency to study investment stocks or flows
between countries, referred to as Foreign Direct Investment (FDI), without
considering a higher resolution. Contrary to this tendency, in this paper, we
focus on investment stocks between regions of Europe, which can be referred to as
Inter-Regional Direct Investment (IRDI). In this sense, as discussed later, some
concepts will be borrowed from the FDI literature and applied in the IRDI context.
The statistical characterization of the network of IRDI stocks in Europe is a
first step towards relating investment flow patterns at a global level to local
and regional dynamics.

FDI is defined as "investment that adds to, deducts from or acquires a lasting
interest in an enterprise operating in an economy other than that of the investor"
the purpose is to have an "effective voice in the management of the enterprise",
as equivalent to holding 10\% or more in the foreign enterprise
\cite{ONS-UNCTAD:FDI_def}. Foreign affiliates are made up of subsidiaries,
associates, and branches. Subsidiaries are majority- or wholly-owned by the parent
companies. Associates are companies in which the investing firm participates in
the management but does not exercise control. Branches are permanent
establishments set up by the parent company in which there is no equity share
capital apart from that of the parent. For associates and subsidiaries, FDI flows
consist of the net sales of shares and loans (including non-cash acquisitions made
against equipment, manufacturing rights, etc.) to the parent company plus the
parent firm's share of the affiliate's reinvested earnings plus total net
intra-company loans (short- and long-term) provided by the parent company. For
branches, FDI flows consist of the increase in reinvested earnings plus the net
increase in funds received from the foreign direct investor. FDI flows with a
negative sign (reverse flows) indicate that at least one of the components in the
above definition is negative and not offset by positive amounts of the remaining
components.  However, the magnitude of FDI flow can also be measured as number of
jobs created or increased. Contribution to more favorable employment status in the
host country is critical when evaluating FDI, especially for countries that are
battling high unemployment rates or want to increase the quality of their
workforce. When quantifying FDI, "flow" (function of time) and "stock"
(cross-sectional/cumulative) are measured and headquarter location of
the investor plays a critical role
\cite{Jakobsen/Onsager:04:HeadOfficeLocation}.

Due to standard legal, institutional and policy attributes that are common across a
given country it indeed makes sense to focus on FDI. However, nowadays firms may
perceive some regions in another country as more similar than regions within
national borders. This might be revealing that the process of European integration
has reduced the national specificities perceived by multinationals and that
regions now are competing to attract FDIs more across than within countries.
Therefore it is important to gain insight on investment flows at a higher spatial
resolution. In this paper we define Inter-Regional Direct
Investment (IRDI) flow in analogy to FDI, as the flow between two administratively
separated regions irrespective of their countries.

\subsection{Relevance of FDI/IRDI for the Global Economy}
Just like FDI, the study of IRDI has prominent policy making implications, in
particular for governing bodies trying to tackle economic growth and employment
creation \cite{IFC:Policies}. There are certain general factors that consistently
determine which countries/regions attract the most investment
\cite{Borensztein/DeGregorio/Lee:98}. In particular, investors cite the following:
market size and growth prospects of the host, wage-adjusted productivity of labor,
the availability of infrastructures, reasonable levels of taxation and the overall
stability of the tax regime. Recent crises have magnified perceptions of
regulatory risks and greater attention is now being focused on the legal framework
and the rule of law. Thus the decision process in investment is multi-factorial
whereas the success or a higher productivity of the investment holds only when the
host country/region has a minimum threshold stock of human capital. Thus,
investment contributes to economic growth only when a sufficient absorptive
capability of the advanced technologies is available in the host economy
\cite{Loungani/Razin}. Promotional efforts to attract investment have become the
focal point of competition among developed and developing countries. This
competition is maintained even when countries are pursuing economic integration at
another level. And it also extends to the sub-national level, with different
regional authorities pursuing their own strategies and assembling their own basket
of incentives to attract new investments. While some see countries lowering
standards to attract FDI in a "race to the bottom," others praise FDI for raising
standards and welfare in recipient countries.  The targets for these promotional
efforts are dominantly the major players of FDI, namely the Transnational
Corporations (TNC's) who on the other end push for newer markets (in the last
decades the global trend of privatization has been a very important mean towards
this end). Public interest driven policies meanwhile continue to serve as the
balancing force, and sustainable development concerns are dealt with accordingly.
The EU has attracted over 40\% of total world flows of FDI in the 1990s, becoming
the largest recipient of multinational activity: multinationals account for a
growing share of gross fixed capital formation in Europe (from 6\% in 1990 to over
50\% in 2000). However, this increasing inflow of FDI in Europe has not been
equally distributed across countries and regions
\cite{Basile/Castellani/Zanfei:03:LocationChoiceMultinat}.

Now, while statistics are mostly regarded at the country-to-country scale,
investments are actually made in specific regions with geographic features, local
administration constraints, and cultural profile. There seem to be a substantial
gap to fill in understanding what is the role of region-to-region investment flow
in the global economy. For this reason we hereby propose to define the IRDI stock
network and we investigate some of its statistical properties as a first stage of
a more comprehensive study to be continued in the future. In section
\ref{sec:Methods} we describe the data set analysed and discuss some
methodological issues. In section \ref{sec:Results} we present and discuss the
results of our analysis. In section \ref{sec:Conclusions} we draw the conclusions
and list some possible extensions of the present work.

\section{Data Sets and Methods}\label{sec:Methods}

In this section we first describe the content of the firm database we used for
our analysis. We then introduce the quantities we have measured on the data set
and include some methodological remarks.

For the firm information we used data collected in December 2004 from the Amadeus
database of Bureau Van Dijk. Access to the on-line service of Amadeus is costly
and was kindly granted by Prof. Delli Gatti of Universit\`{a}  Cattolica di
Milano. The database provides firm address, financial profile, number of
employees, industrial classification, names of shareholders and board of directors
of virtually all firms in Europe.

However data are delivered in files of limited size and we were forced to restrict
the download to a selection of the ensemble of firms. We have chosen to select
the firms with number of employees larger or equal to 100. Also, a small
percentage of around 1\% of firms was lost in this selection because the
number of employees was not available. The original data set consists of 181.945
 firms from 39 European countries uniquely identified by their Bureau Van Dijk identification
number (BVDID). For a given firm, shareholders can be individuals/families,
governments or other institutions that are not listed as firms. Even when they are
firms they may have only indirect ownership through intermediate firms. Moreover,
they may have less than 100 employees and therefore location and profile for them
is not available in our data set. The set of firms which are involved in ownership
links with other firms includes 47.621 firms. We have chosen to restrict our
analysis to the direct ownership network of firms with more than 100 employees,
implying that we have the financial profile and geographical location of all the
firms involved in this set of ownership links. This is the set of data we use for
the network analysis and it includes 29.314 firms and 22.174 links.

The selection of a subset of the ownership links induces of course an
underestimation of the total hosted investment stock. Still, investigating how
investment size among this set of firms is distributed in the network and
among geographical regions is a very interesting point to address.

\subsection{Defining the Quantities of Interest}\label{sec:Methods:Quantities}

For each firm $i$ we consider the following quantities: the activity $a_{i}$
measured as number of employees in the firm, the shares $w_{ij}$ of firm $i$ owned
by any other firm $j$, and the headquarter region $R_{i}$ of firm $i$. The number
of employees is one of the standard quantities used to measure firm size
\cite{Axtell:01:Science}, and in the following we will measure also investment
stocks in terms of number of employees. There are of course other possible
measures of investment stocks, based on capital rather than on human resources,
but from the point of view of labour market and economic impact at a local scale,
it is relevant to have an estimate of how many employees of a firm or a region
depend on the investment coming from outside.

\begin{figure}[htb]
\centerline{ \includegraphics[width=1\textwidth]{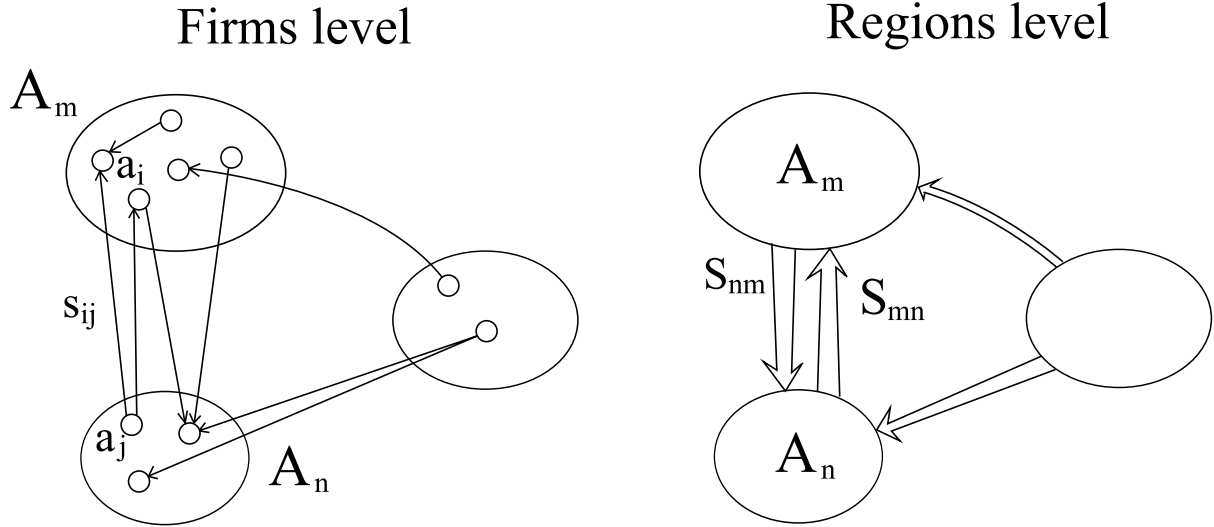}}
\caption{Diagram illustrating how the (right) region network $G_R$ is built up from the (left) firm
network $G_F$. \label{diag_firm_reg} }
\end{figure}

Shares are usually defined as a percentage, but it is more convenient to define
$w_{ij}$ as a fraction of ownership and, therefore, as a real number in $[0,1]$.
Not all shares are necessarily held by entities external to the
firm. Moreover some entities are not firms and we only look at shares held by
firms, as discussed above. Therefore it holds:

\begin{equation} \label{eq:s_restriction}
  \sum_{j\neq i} w_{ij} \leq 1
\end{equation}
If we take the number of employees as a measure of activity of a firm, it is
natural to compute the quantity $s_{ij}$:
\begin{equation} \label{eq:w_ij}
  s_{ij} = w_{ij} a_{i}
\end{equation}
that represents the investment stock of firm $i$ held by firm $j$.

We can define the \textit{firm network} as a graph $G_F=(V_F,E_F)$ Figure \ref{diag_firm_reg}(left) where $V_F$ is
a set of nodes representing firms and $E_F$ is a set of directed edges between
nodes. An edge $(i,j)$ represents the fact that firm $j$ owns shares of firm $i$. The
order of the pair in this notation is natural for the layout of most databases of
firms and we adopt it. However, keep in mind, that ownership
and investment have opposite directions. In fact, it is more natural to define
\textit{in-degree} and \textit{out-degree} of connectivity with respect to
investments rather than to ownership. We define as \textit{in-degree} $k_{in}$ of
a firm the number of firms investing in $i$ (holding shares of $i$). Similarly we
define as \textit{out-degree} $k_{out}$ of a firm the number of outside firms in
which firm $i$ invests. The \textit{connectivity degree} or simply degree $k$ of a
node is the number of edges entering or departing from that node.

We can associate to each edge $(i,j)$ the normalized weight $w_{ij}$, representing
the amount of shares that firms $j$ owns in $i$. But we can also associate the
absolute weight $s_{ij}$. We then define the \emph{inward investment stock}
$a_{i}^{in}$ of firm $i$ , as the total stock invested in firm $i$ by other
firms and \emph{outward investment stock} $a_{i}^{out}$, as the total stock
invested by firm $i$ in other firms:

\begin{equation} \label{eq:a_in}
  a_{i}^{in} = \sum_{j} s_{ij}  \hspace{1cm} a_{i}^{out} = \sum_{k} s_{ki}
\end{equation}

We can now define analogous quantities aggregated by region. The \textit{activity
of region $m$} is defined as the sum of the activity of the firms with headquarter
in that region.

\begin{equation} \label{eq:A}
  A_{m} = \sum_{i \in m} a_{i}
\end{equation}

In other words, $A_{m}$ is the total number of workers employed by firms of that region.
Keep in mind that we analyse only a subset of all firms and that therefore, the
activity of a region can be much smaller than the number of individuals employed
in that region. The sum of the investments made by firms of region $n$ in firms in
region $m$ is defined as:
\begin{equation} \label{eq:W}
  S_{mn} = \sum_{i \in m, j \in n} s_{ij}
\end{equation}
It can be seen both as the outward investment stock of region $n$ in region
$m$ or as the hosted investment stock in region $m$ coming from region $n$

It is very natural at this point to define the \textit{region network} as a graph
$G_R=(V_R,E_R)$ where nodes represent regions and a directed edge $(m,n)$ from
region $m$ to region $n$ represents the fact that some firms of region $n$ own shares in
some firms of region m. The diagram in figure \ref{diag_firm_reg} illustrates the
procedure of building the network of regions. Small circles represent firms with
their associated values of activity. Edges represent ownership relations. Larger
circles represent regions in which firms have their headquarters. The edges in
firm network among all firms in regions $m$ and $n$ sum up to form an edge between m
and $n$ in the region network. We associate to the edge $(m,n)$ the absolute weight
$S_{mn}$. The degree is defined as for the firm region and in particular in-degree
and out-degree are defined with respect to investments.

The sum of the investments made in firms of a region by firms of any other region
will be called \textit{inward investment stock of region m} (eq.\ref{eq:A_inout}).
In the following we will refer to this quantity also as \textit{hosted investment
stock}. Similarly, the sum of the investments made by the firms of a region in
firms of other regions will be called \textit{outward investment stock of region
m} (eq.\ref{eq:A_inout}).
\begin{equation} \label{eq:A_inout}
  A_{m}^{in} = \sum_{n} S_{mn}   \hspace{1cm} A_{m}^{out} = \sum_{n} S_{nm}
\end{equation}
Both definitions are chosen in analogy with the terms used in the literature about
Foreign Direct Investments. But instead of looking at investments between different
countries, we increase the spatial resolution to the level of regions.

As only the few most important shareholders of each firm are usually listed in the
database, the in-degree of firms is a bias quantity with little meaning for our
purposes. However, when aggregating by region, the in-degree of a region represents
the number of other regions in which the top shareholders of firms in the focal
region have their headquarters. This quantity is not limited a priori and as such it
makes sense to study its distribution.

\subsection{Measuring the Quantities of Interest}\label{sec:Methods:remarks}
Our perspective in this work is to try and relate the microscopic and macroscopic
aspects of the network of investments between firms and between regions. We want
to look at statistical distributions and not at single or average values, the aim
being to try and understand what is the individual tendency that builds up the
macroscopic properties. Therefore we will focus on the probability distributions
of the quantities defined above.

In order to study the probability distribution (pdf) of a variable $x$ it is
useful to plot its \textit{cumulative distribution function} (cdf) defined as
\begin{equation} \label{eq:cdf}
  P_C(\hat{x}) = \int_{x\geq\hat{x}} p(x) d x
\end{equation}
where $x \rightarrow p(x)$ is the probability distribution function. In words, the
cdf gives the fraction of a randomly chosen sample of the variable $x$ that lies
above the value $\hat{x}$. A simple way of constructing $P(x)$ is the following.
Consider the vector $x$ of $N$ real numbers. We rank $x$ in ascending order.
Clearly, now all values are larger or equal to the first data point. So the
probability distribution starts from 1 and decreases. The k-th component of the
vector $x$ has ascending rank $k$ and there are $N-k$ values larger or equal to
$x(k)$. The fraction of data larger or equal to $x(k)$ is $N-k/N$. We therefore
simply plot the pair $(x(k),N-k/N)$ for all $k$. If some values of $x$ are
repeated and in particular if $x$ is a discrete variable, then the plot will
display 'stairs'. In this case it is preferable to count the fraction $P$ of data
that are larger or equal to each value $x$ and then to plot $P$ versus $x$.

If the distribution of the variable $x$ is a power law with exponent $-\gamma$,
then its cumulative distribution is still a power law with exponent $-\gamma+1$.
In $\log-\log$ scale they appears as straight lines with different slope:
\begin{eqnarray}
   p(x)= c_1 x^{-\gamma}\\
   P(x)= c_2 x^{-\gamma+1}
\end{eqnarray}
where $c_1,c_2$ are normalization factors. If instead the distribution of the
variable $x$ is log normal with coefficients $\mu$ and $\sigma$ (eq.
\ref{eq:lognormal}), then it appears as a quadratic curve in $\log-\log$ scale.
\begin{eqnarray}\label{eq:lognormal}
   p(x)= \frac{1}{x\sqrt{2\pi\sigma^2}} exp (- \frac{(\log x - \mu)^2}{2\sigma^2}\\
   \log(p(x)) \sim -\log(x) - \frac{(\log x - \mu)^2}{2\sigma^2}
\end{eqnarray}
However, its cumulative distribution does not have an analytical expression.

It is usually more accurate and safe to estimate the exponent from the cdf rather
than from the pdf, because of the fluctuations in the frequency of high values of
$x$. This is usually done by fitting the cdf in $\log-\log$ scale with a line and
computing the slope. However, it has been recently remarked that this method
introduces a systematic bias \cite{Goldstein/Morris/Yen:04}. An alternative method
that doesn't make use of a graphic fit is described in \cite{Newman:05:PLReview}.
The formula for the exponent is
\begin{eqnarray}
   \gamma=1+N\left[\sum_i \log\frac{x_i}{x_{\mathrm{min}}}\right]^{-1}
\end{eqnarray}
where $x_{\mathrm{min}}$ is the lower limit of the range of data following the
power law. Confidence intervals for $\gamma$ can be computed with the standard
bootstrap technique.

When a relation holds between two variables, then their respective probability
distributions are also related through the equation:
\begin{eqnarray}
  p(y)=p(f(x))=p(x)\frac{df}{dx}
\end{eqnarray}
An important consequence is that if, $x$ is power law distributed and $y$ scales as
a power law of $x$, then $y$ is also power law distributed, and a relation holds
for all the exponents involved (power law distributions are closed with respect to
the operation of power law rescaling). If $x$ is log normal distributed, and $y$
scales as a power law of $x$, then the distribution of $y$ converges to a log
normal function for large $y$. This observation implies that, if two variables are
both power law distributed, and we suspect that a relation might hold between them,
such relation could be a power law scaling, therefore we have to be careful
while checking for possible correlations.

In fact, in the study of the correlation between quantities, related to firms and
regions, that span several order of magnitude and have to be studied in $\log-\log$
scale, we proceed as follows. Consider the variables $(X,Y)$ for instance. We
compute the $\log{10}$ of both variables and produce a scatter plot of
$(\tilde{X},\tilde{Y})=(\log{10} X,\log{10} Y)$. We then divide the X axes in $k$ bins
of equal size. For each bin centered in the value $x_i$, we compute the mean $y_i$
and the standard deviation $\sigma_{y_i}$ of the values of $Y$ for which the corresponding
abcissa falls in the bin $k$. We obtain a new set of data points
$(x,y)$ that allows to better display the trend of the original data $(X,Y)$. A
linear fit is then performed on the $(x,y)$ data points, and values of slope,
intercept and correlation coefficient are computed. We remind that a linear
relationship between $x$ and $y$ implies taking the exponential, a power law
relationship between the original variables:
\begin{eqnarray}
 \tilde{Y}= m \cdot \tilde{X}+q\\
 \log{10}(Y)=m \cdot \log{10}(X) + q \\
 Y=C \cdot X^m
\end{eqnarray}
where $C=10^q$. Of course, while the operation of binning and averaging over bins
is a standard procedure one should be careful in this case, as such operation does
not commute with the operation of taking the exponential. However, we are not aware
of any documented bias introduced by this procedure and if the fit is reasonably good
we conclude that $Y$ is scaling as a power law of $X$ and we take $m$ as the
exponent of the scaling law.

\section{Analysis at the level of firms}\label{sec:Results}

We first report the cumulative frequency distribution of activity and investment
stock of firms. We also investigate how investment stock scale with firm activity
and with firm connectivity degree. We then report the analogous results for
regions where activity, investment stock and connectivity degree of regions are
defined as in section \ref{sec:Methods:Quantities}. Similarly, we investigate how
investment stock scale with region activity and region connectivity degree.

\subsection{Firms. Distributions of activity,investment and connectivity degree}\label{sec:Results:cdf_comp}

\begin{figure}[th]
\centerline{ \includegraphics[width=0.47\textwidth]{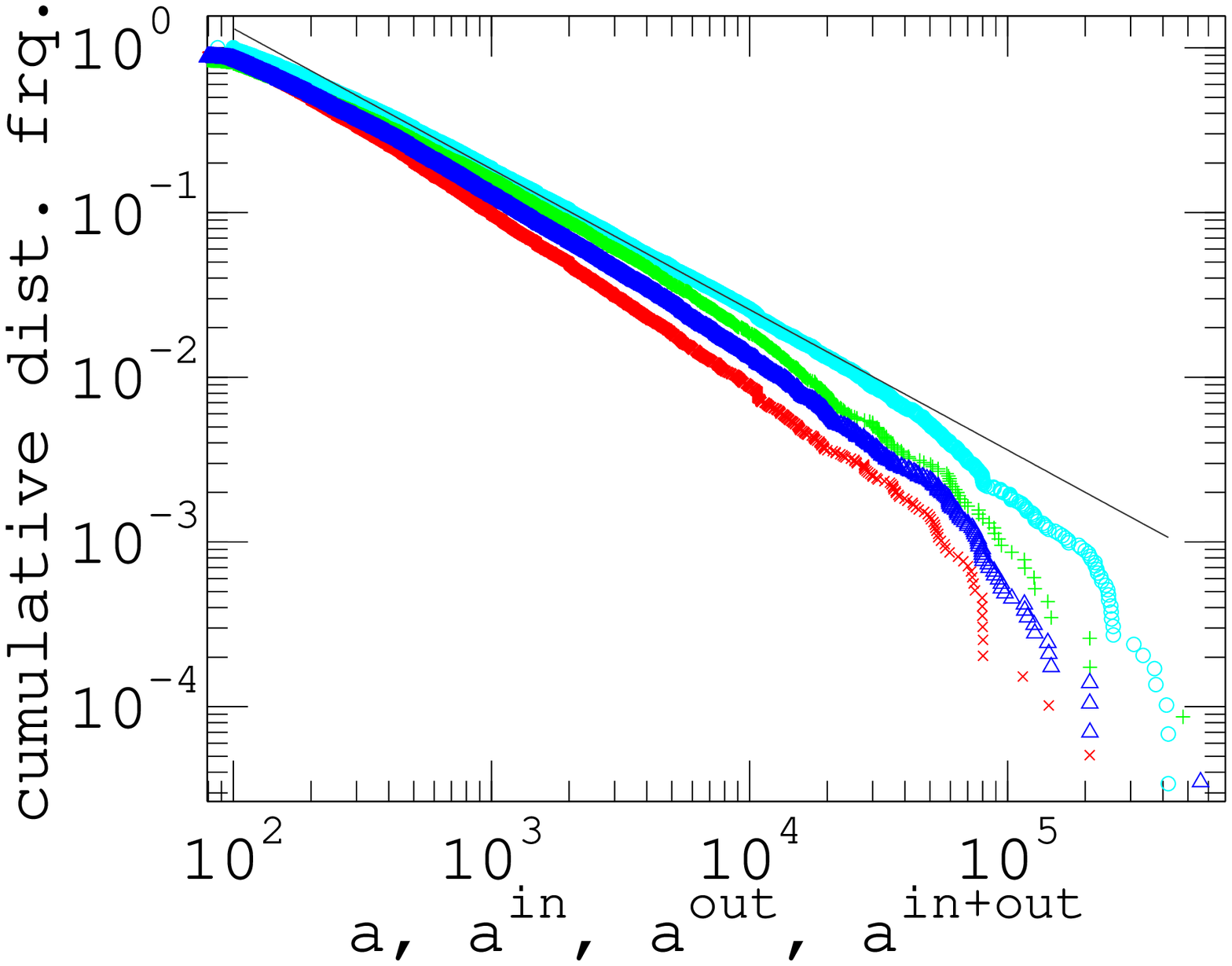}
\hfill
\includegraphics[width=0.47\textwidth]{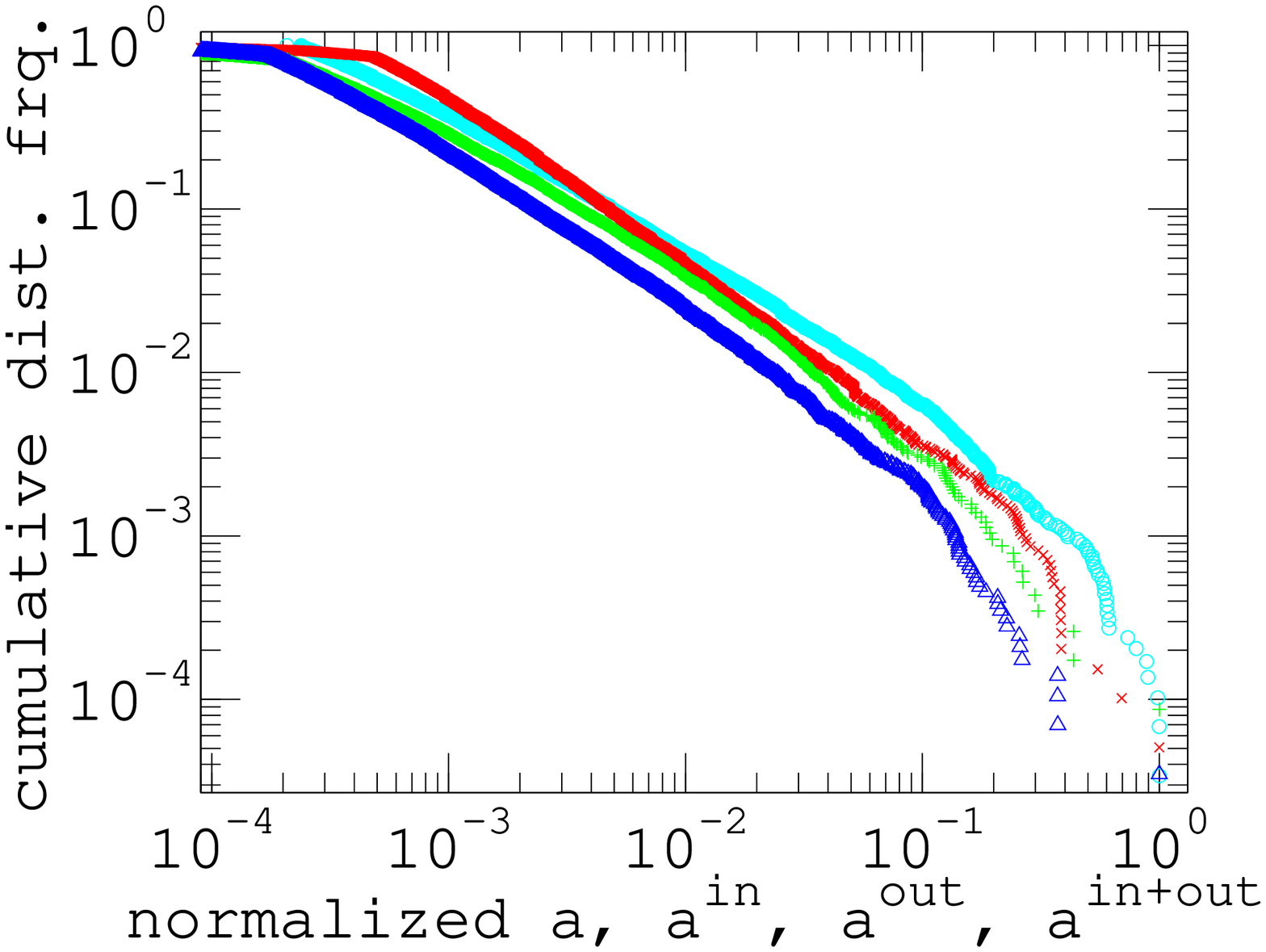}}
\caption{Firms. Cumulative distribution of: (left)  activity $a$ (o); hosted
investment stock $a^{in}$ ($\times$); outward investment stock $a^{out}$ (+); total
investment stock $a^{in+out}$ ($\Delta$) and (right) normalized values with respect
to the maximum value\label{fig_Comp_cdf_flux}}
\end{figure}

In figure \ref{fig_Comp_cdf_flux} we report the cumulative distribution (cdf) of
activity and investment stock of firms computed from our data set. The onset at
the value 100 for the activity is simply due to the restriction of the data set to
firms with more than 100 employees. Investment stock can of course take smaller
values as it is measured as a fraction of the number of employees per firm. The
cumulative distributions display a linear decay over 3 decades or more. However
some 'bumps' deviating from linearity are visible. The curves can either be fitted
with $\log$ normal distributions with very large standard deviation or can be
reasonably fitted with a power law. In both cases the meaning is the same: the
probability of finding large firms is decreasing approximately as the exponent of
the power law fitting the curve. Computing the exponent from the linear fit is
known to introduce bias \cite{Goldstein/Morris/Yen:04}, hence following
a known procedure \cite{Newman:05:PLReview} the values of the exponents and confidence intervals
were computed as described in \ref{sec:Methods}. The values of the exponents and
their confidence interval are reported in table \ref{tab:firm_cdf}.

\begin{table}[h]
\caption{Firms. Power law fit values for Cumulative Distributions.
\label{tab:firm_cdf}} {\begin{tabular}{@{}ccc@{}} \toprule
    Data    &    $\gamma$   &  $\sigma_{\gamma}$   \\ \colrule
 $a$          &   1.7829   &     0.0038            \\
 $a^{in}$     &   1.9307   &     0.0064            \\
 $a^{out}$    &   1.7684   &     0.0061            \\
 $a^{in+out}$ &   1.8480   &     0.0047            \\
 $k^{out}$    &   3.849    &     0.044            \\ \botrule
\end{tabular}}
\end{table}

Our finding concerning the activity is in line with what is generally known in the
literature for firm size distributions of many countries and historical epochs. It
implies that firm activity is very heterogeneous and that roughly speaking, very
large values and very small values of activity are much more frequent that in
normal distributions. We remind the reader that the data set analysed includes
only firms involved in an ownership relationship in Europe and not all firms
indiscriminately. However, the value of exponent $\gamma$ is not far from results
obatined in previous studies. For instance Axtell reports 2.056 for the US
 firm activity distribution \cite{Axtell:01:Science}. Fujiwara et al. report 1.995 for UK based on data from
Amadeus Database \cite{Fujiwara/DiGuilmi/Aoyama/Gallegati:03:ParetoZipf}. On the
other hand, the fact that investment stock distribution is also a power law is to
our knowledge a novel result. In particular, the values of the exponent $\gamma$
for activity and outward investment stock are very close.

This finding might suggest that a scaling relation holds between the two variables
(as discussed in section \ref{sec:Methods:remarks}). However, this hypothesis had
not been empirically verified so far in the literature and we don't know a priori
to what extent it holds. We will then investigate its validity in the next
section.

\begin{figure}[th]
\begin{center}
\centerline{ \includegraphics[width=0.47\textwidth]{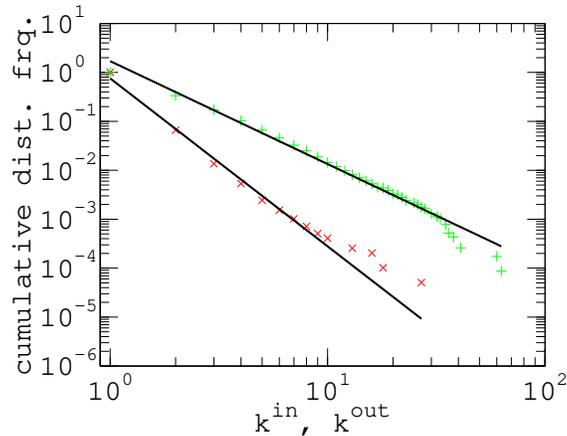}}
\caption{Firms. Cumulative distribution of connectivity degree. In-degree
($\times$) , out-degree (+) \label{fig_Comp_cdf_deg}}
\end{center}
\end{figure}

As is usual in the study of networks, we report the cumulative distribution functions
(cdf) of the degree of connectivity of firms (figure \ref{fig_Comp_cdf_deg} ). We
distinguish between (total) degree, in-degree and out-degree as defined in
\ref{sec:Methods:Quantities}. The distributions span a short range of less than
two decades. The out-degree display a linear decay in $\log-\log$ scale that can be
fitted by a power law with exponent 3.85, which is a little larger than the
typical values observed in many empirical complex networks (typically in the range [1.8 3]).
The result implies that the number of connections between firms is moderately
heterogeneous and decreases slower than exponentially. The curve for the in-degree
instead is not meaningful as mentioned in section \ref{sec:Methods:remarks} and
simply shows how many records of shareholders are available per firm. The linear
fit was performed just for sake of completeness.

However, it is important to remark that the connectivity out-degree represents the
number of ownership relations in which the firm is involved, regardless of the
size of shares involved in each relation. Because the degree doesn't take into
account the size of the shares, a large out-degree doesn't mean that a firms really
controls a lot of other firms. Alternative quantities are needed to characterize
the ownership concentration such as those introduced in
\cite{Battiston:04:InnerStructCapitNets} and they will be applied to the present
data set in a future study.

\subsection{Firms. Correlations among activity, investment and connectivity degree}\label{sec:Results:corr_comp}
\label{sec:Results:correl_comp} 
\begin{figure}[th]
    \begin{center}
      \includegraphics[width=0.47\textwidth]{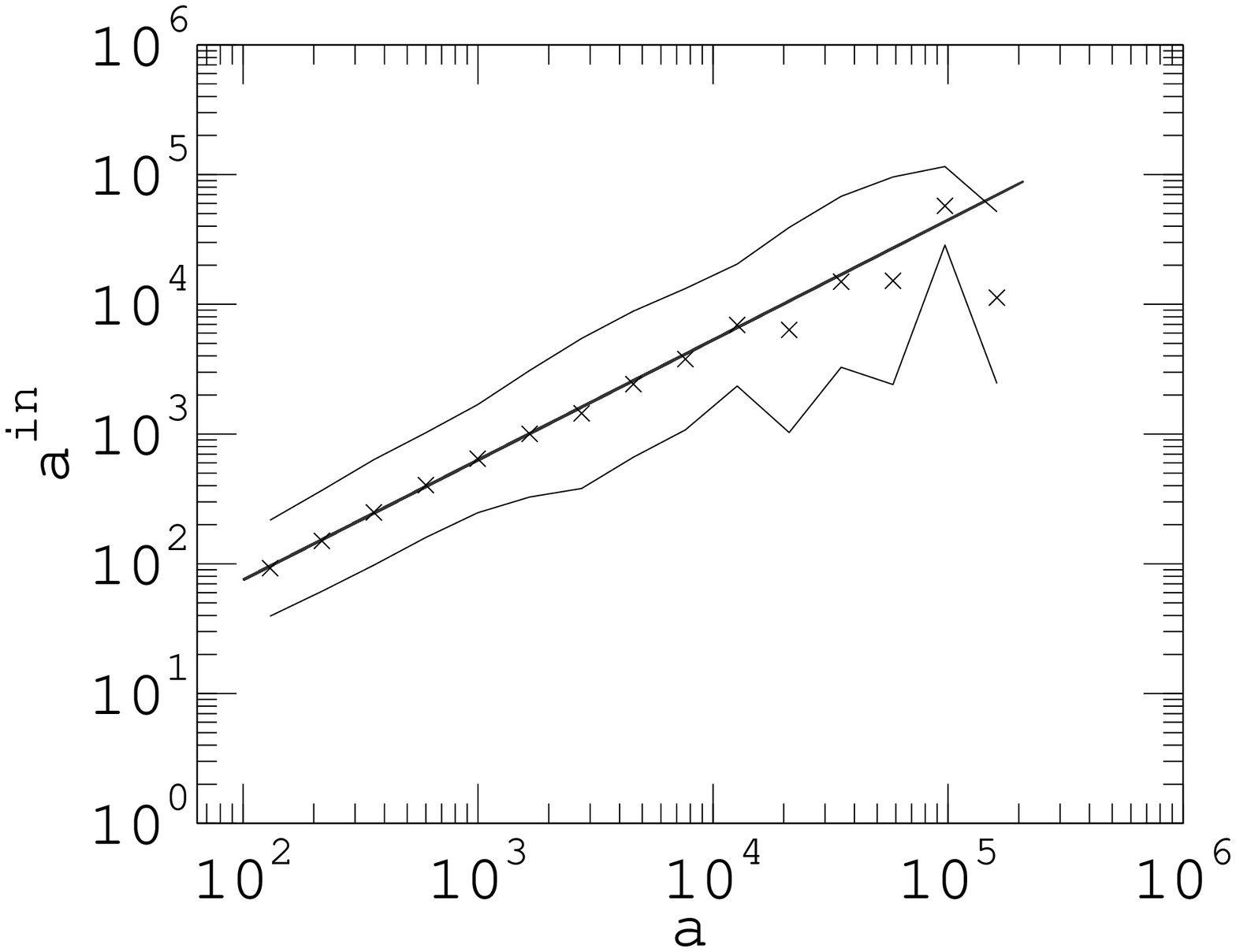}
      \hfill
      \includegraphics[width=0.47\textwidth]{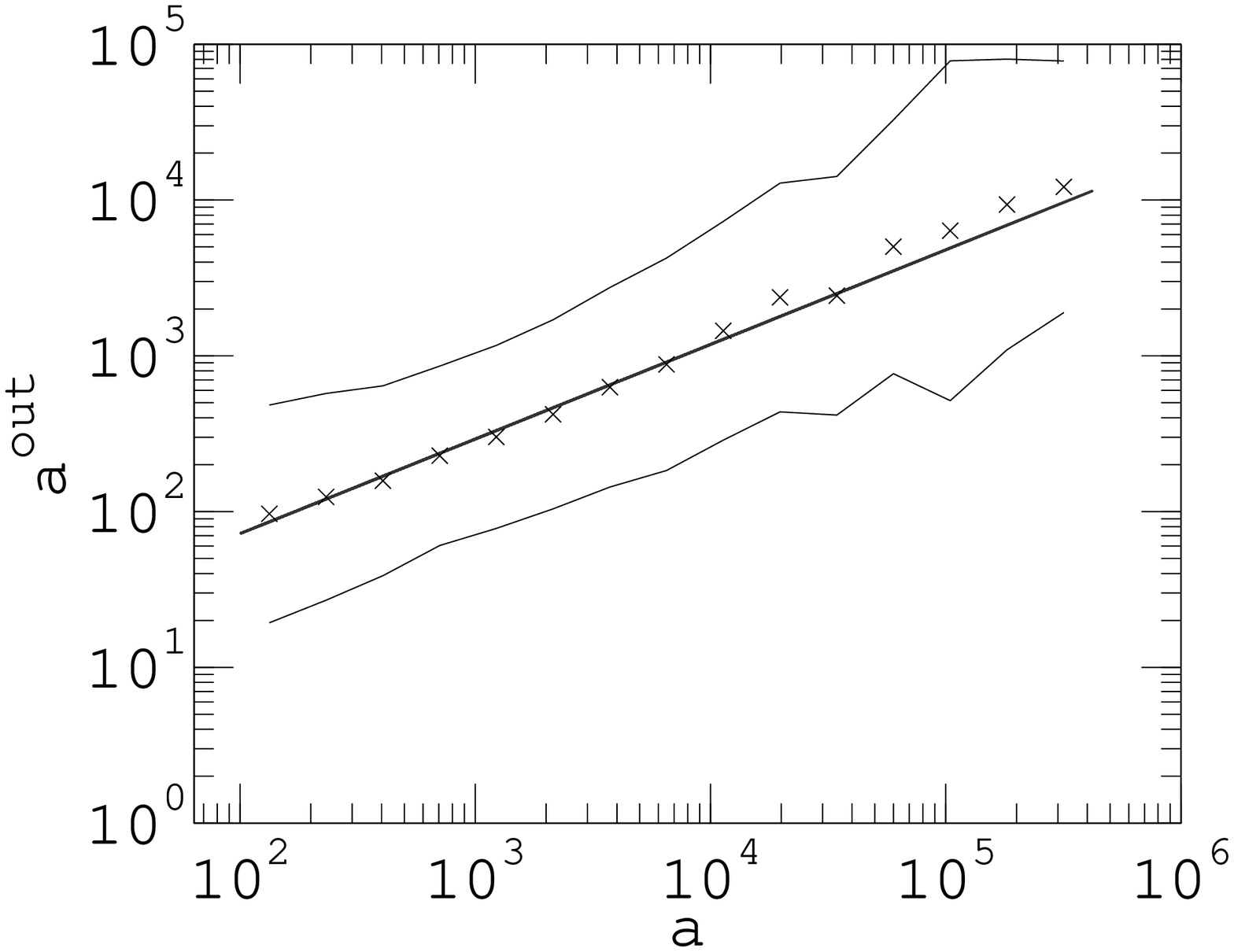}
      \caption{Firms. Plot of (left) hosted invested stock $a^{in}$
       and (right) outward invested stock $a^{out}$ of firms versus their
       activity $a$ in $\log-\log$ scale. Data are binned.
       Mean $\pm$ standard deviation of the values in each bin are plotted as the continuous lines.
      \label{fig_co_corr_in_out}}
    \end{center}
\end{figure}

\begin{figure}[h]
    \begin{center}
      \includegraphics[width=0.47\textwidth]{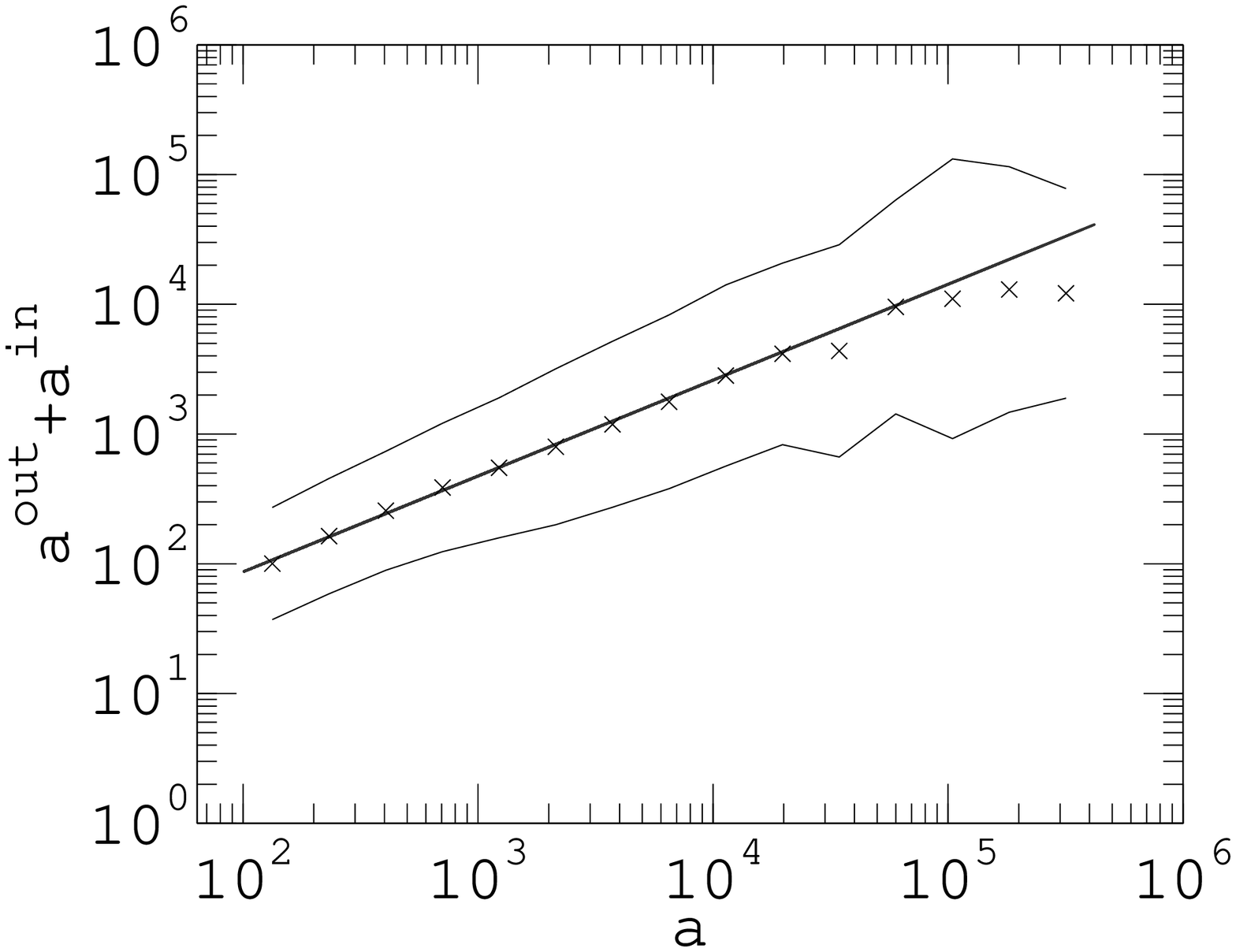}
      \hfill
      \includegraphics[width=0.47\textwidth]{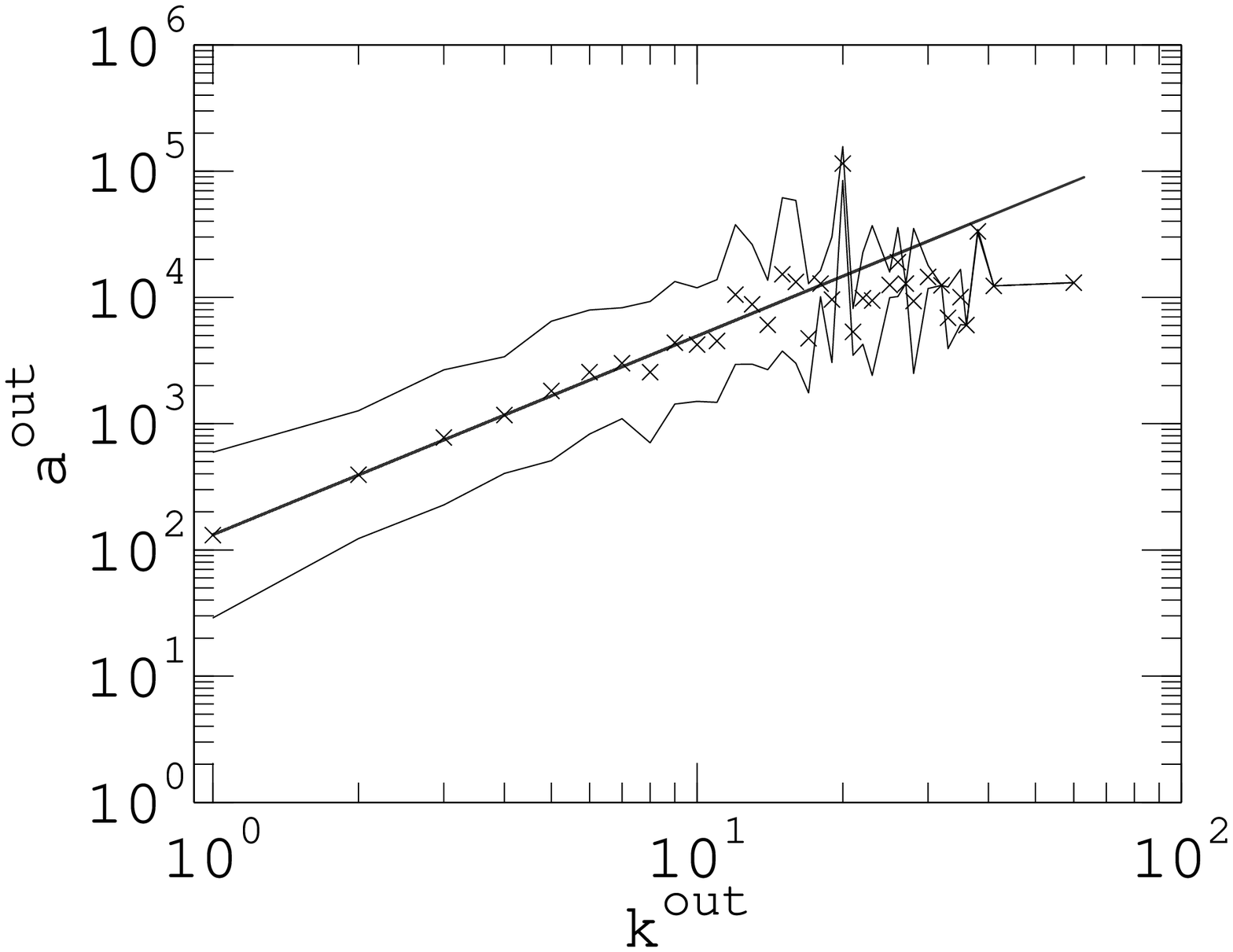}
      \caption{Firms. (left) Plot of total invested stock $a^{in+out}$ of
      firms versus their activity $a$. (right) Plot of connectivity out-degree $k^{out}$ of
      firms versus their hosted investment stock $a^{out}$ in $\log-\log$ scale.
      Plot Data are binned.
      Mean $\pm$ standard deviation of the values in each bin are plotted as continuous lines.
      \label{fig_co_corr_inout_kout}}
      \end{center}
\end{figure}

In order to understand why investment of firms are
power law distributed and why distributions of outward investment and activity
have exponents very close to each other we investigate the correlations between
activity, investment and connectivity out-degree. The plots in figures
\ref{fig_co_corr_in_out} - \ref{fig_co_corr_inout_kout} are produced by
taking the $\log10$ of the quantities and binning the data on the x axes as
described in section \ref{sec:Methods:remarks}.

Table \ref{tab:firm_corr} reports the value of slope and correlation coefficients
for the linear fit of the binned data. The correlation coefficients are all quite
close to 1 so, with the caveat mentioned in section \ref{sec:Methods:remarks}, we
conclude that the data indicate the existence of scaling laws between investment
and activity and between out-degree and activity.

 \begin{table}[h]
 \caption{Firms. Scaling of investments versus activity.
 Coefficients of linear fit and correlation coefficients.\label{tab:firm_corr}}
 {\begin{tabular}{@{}cccc@{}} \toprule
    Yaxis     &   $m$      &   $q$     &    corr.coef.    \\ \colrule
 $a^{in}$     &   0.925   &  0.026   &    0.974        \\
 $a^{out}$    &   0.607   &  0.644   &    0.997        \\
 $a^{in+out}$ &   0.739   &  0.460   &    0.992        \\ \botrule
 \end{tabular}}
 \end{table}

We notice that the exponent $0.925$ for the scaling of hosted investment versus
activity is smaller than, but still close to 1. This means that firms tend to host
investment in amount almost just proportional to their activity. This is not
surprising if we consider that all firms in the data set analysed, are owned to some
extent by some other firm in the data set and in many cases the share is
large and close to 1. A deeper understanding would require an investigation of the
statistics of the ownership concentration and will be carried out in a future
work. On the other hand the exponent $0.61$ for the scaling of outward investment
versus activity implies that although more active firms tend to invest more, the
investment increases less than linearly as a function of the activity (sub-linear
increase). This means that very large and active firms invest less, in proportion,
with respect to smaller ones. Such a finding becomes quite important if an
institution in charge of attracting investments from firms is trying to estimate
the expected investment of firms based on their activity.

Finally, the exponent $1.57$ for the scaling of activity versus out-degree is
quite interesting. It implies that the larger the firm the larger the number of
investments but the increase is sub-linear. Interestingly, the value of the
exponent is not far from the values found for the scaling law between invested
volume and degree in some stock markets: 1.1 for Nasdaq,  1.43 for NYSE, 1.59 for
MIB \cite{Garlaschelli/Battiston/al:05:SFTopo}. In that case the invested volume is
exactly the analogous quantity to the outward investment. Moreover, the scaling law
resembles the one observed inthe context of air traffic networks for a quantity
analogous to the outward investment which has been recently introduced as node strength
\cite{Barrat/Barth/Vesp:05:SpatConstrInAirNet}. In that case the exponent found is
$1.7$.  These findings might support the idea that a universal scaling law for node
strength holds in complex networks where the weight plays a crucial role. In a
future work we will investigate possible network formation models leading to the
emergence of such scaling law in economic networks.

\vspace{.2cm}
\begin{table}[h]
 \caption{Firms. Scaling of investments versus connectivity degree.
 Coefficients of linear fit and correlation coefficients \label{tab:firm:kcorr}}
 {\begin{tabular}{@{}cccc@{}} \toprule
  Yaxis       &   $m$      &   $q$     &    corr.coef.    \\ \colrule
 $a^{out}$    &   1.574   &  2.120   &    0.855        \\ \botrule
 \end{tabular}}
 \end{table}

\section{Analysis at the Level of Regions}\label{sec:Results:Regions}
\subsection{Regions. Distributions of activity, investment and connectivity degree}\label{sec:Results:cdf_reg}

\begin{figure}[th]
\centerline{ \includegraphics[width=0.47\textwidth]{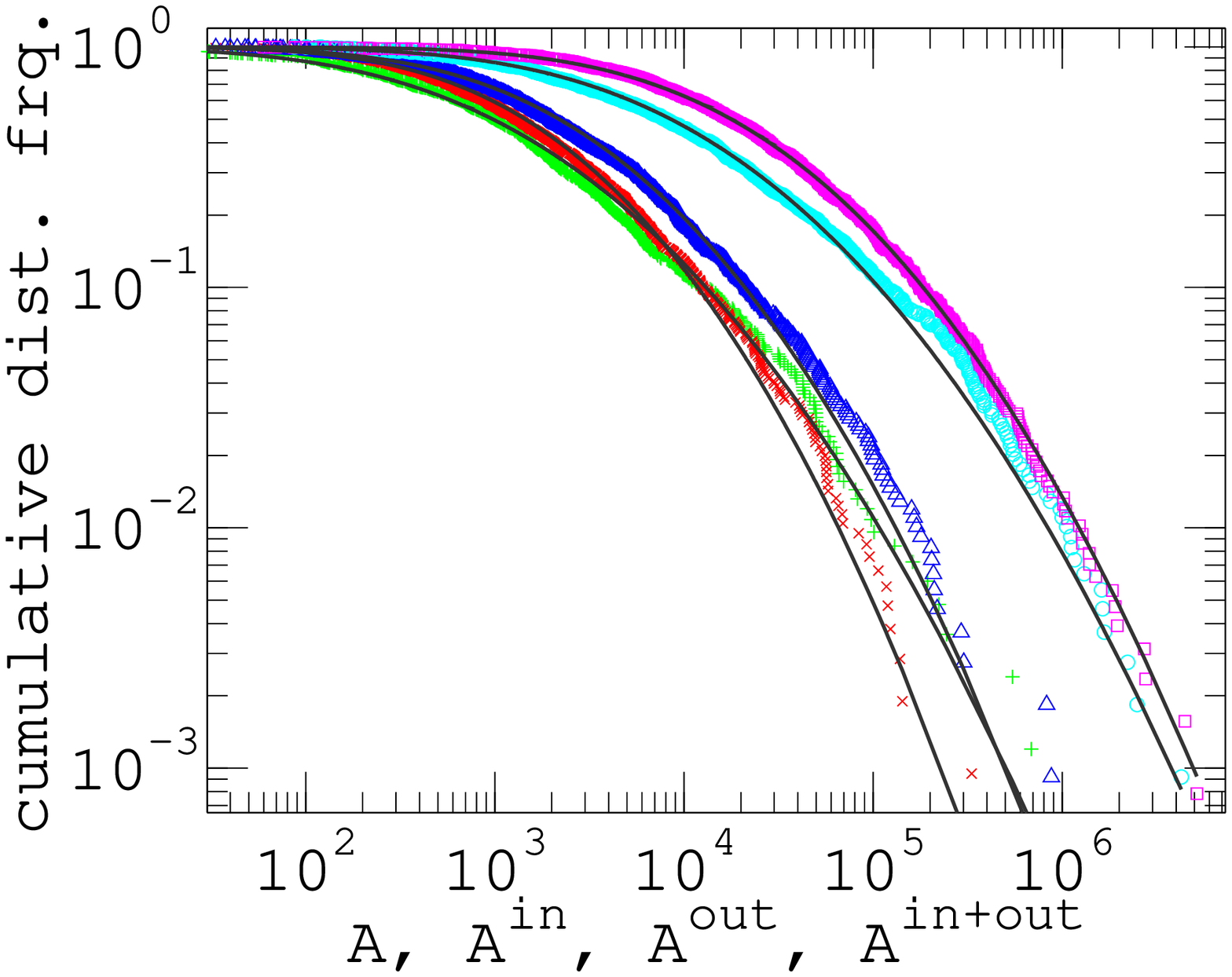}
\hfill
\includegraphics[width=0.47\textwidth]{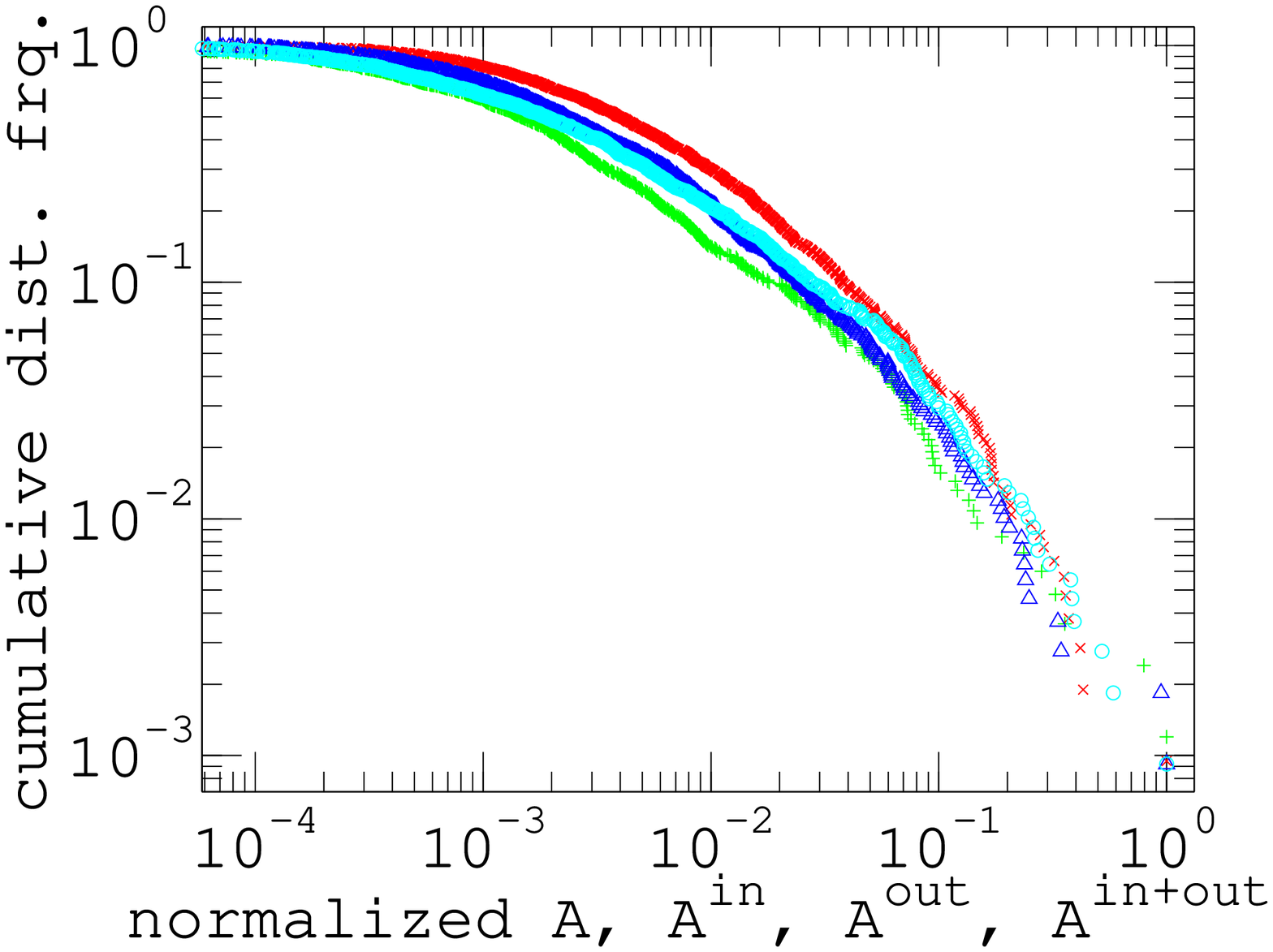}}
\caption{Regions. Cumulative distribution of: (left) activity $A$ (for all firm profiles) ($\square$); activity $A$ (o); hosted investment
stock $A^{in}$ ($\times$); outward investment stock $A^{out}$ (+); total
investment stock $A^{in+out}$ ($\Delta$) and (right) the same quantities normalized with respect to the maximum value. Zoom.
\label{fig_reg_cdf_flux}}
\end{figure}

In figure \ref{fig_reg_cdf_flux} (left) we report the cumulative distribution of activity
$A$ and inward/outward/total investment stock $A^{in}$, $A^{out}$, $A^{in+out}$ for
regions. These quantities were computed from the data set as described in section
\ref{sec:Methods:Quantities}. As done for the case of firms, in order to check to what extent they overlap
we also normalized such quantities and we report their cumulative distribution in figure
\ref{fig_reg_cdf_flux} (right) . As it
can be seen, the overlap is only partial. This may suggest that a non linear
scaling holds between the variables and this hypothesis will be investigated in
the next section.

It must be remarked that the way regions are defined within a country surely
depends on the country's administrative system. To give an example, the regions
provided in the data set are at the level of 'provincia' for Italy and
'department' for France which have comparable surface on average.

\begin{table}[h]
\caption{Regions. Log normal fit values for Cumulative
Distributions.\label{tab:reg_cdf}} {\begin{tabular}{@{}ccccc@{}} \toprule
    Data             &  $\mu$   &  $\Delta\mu$      &   $\sigma$ &  $\Delta\sigma$   \\ \colrule
 $A$                 &   9.06   &     0.12          &    1.97    &      0.08       \\
 $A_{all dataset}$   &   9.79   &     0.10          &    1.82    &      0.07       \\
 $A^{in}$            &   7.28   &     0.10          &    1.64    &      0.07       \\
 $A^{out}$           &   6.89   &     0.14          &    2.02    &      0.10       \\
 $A^{in+out}$        &   7.69   &     0.10          &    1.76    &      0.07       \\
 $K$                 &   1.98   &     0.07          &    1.18    &      0.05       \\
 $K^{in}$            &   1.50   &     0.06          &    1.00    &      0.04       \\
 $K^{out}$           &   1.52   &     0.08          &    1.16    &      0.06       \\ \botrule
\end{tabular}}
\end{table}

\begin{table}[h]
\caption{Regions. Power law fit values for Cumulative
Distributions.\label{tab:reg_powl_cdf}} {\begin{tabular}{@{}ccc@{}} \toprule
    Data           &    $\gamma$   &  $\sigma_{\gamma}$   \\ \colrule
 $A$                 &   2.29   &     0.14            \\
 $A_{all dataset}$   &   2.23   &     0.09            \\
 $A^{in}$            &   2.58   &     0.19            \\
 $A^{out}$           &   2.40   &     0.20            \\
 $A^{in+out}$        &   2.21   &     0.11            \\
 $K$                 &   2.53   &     0.09            \\
 $K^{in}$            &   3.00   &     0.15            \\
 $K^{out}$           &   2.40   &     0.10            \\ \botrule
\end{tabular}}
\end{table}

Differently from the case of the firms, the distributions do not display a linear
decay in $\log-\log$ scale but rather a quadratic one which we tried to fit with
log-normal distributions. The fit with log normal is quite good
although a discrepancy can be observed in the right tail. On the other hand, only the
very last portion of the distribution could be fitted with a power law. Values for
the coefficients of the log normal fit are given in table \ref{tab:reg_cdf}. Just
for sake of comparison we also report the exponent of the power law fit of the
rightmost tail (table \ref{tab:reg_powl_cdf})

At a first sight the finding that the distributions above are log normal is
puzzling, as one may expect that region activity and investment stocks also scale
with a power law. However, the following remark is relevant at this point. Cities
ranges from small villages of few inhabitants to metropolis of 10-20 millions.
Firms also range from one-person enterprises to multinationals with few hundreds
thousands employees. On the other hand, while the position of the boundaries of a
region surely is the result of the historical process, the surface and the amount
of population within a region is probably limited by administration tasks
constraints, in the sense that if the administrative load becomes too heavy, the
region is split into two. For instance within a same country there are no regions
which are orders of magnitude larger in surface than other regions. On the other
hand, it is known that activity of countries measured by Gross Domestic Product is
power law distributed with exponent 1 \cite{Garlaschelli/Loffredo:04:WTW}. So it
appears that regions are clearly less heterogeneous with respect to firms and
countries. However before trying to draw some implications for economic
development policies, it would be interesting to normalize the activity of the
regions by the regions surface and/or active population. Unfortunately these data
were not available during this work.

\begin{figure}[th]
\centerline{ \includegraphics[width=0.47\textwidth]{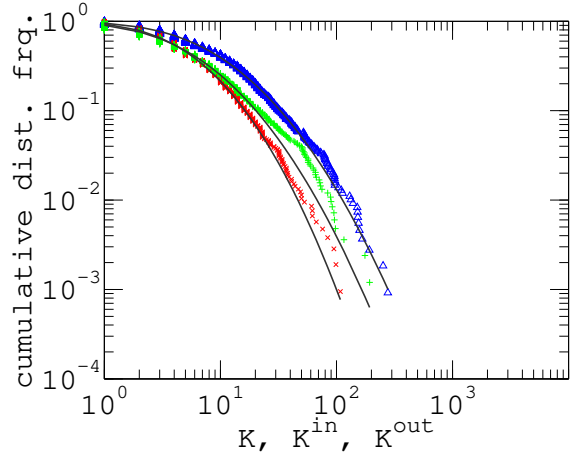}}
\caption{Network of Regions. Cumulative distribution of connectivity degree.
In-degree ($\times$) , out-degree (+), total degree ($\Delta$) \label{fig_reg_cdf_deg}}
\end{figure}

In figure \ref{fig_reg_cdf_deg} we report the cdf of the connectivity degree for
regions. All curves are clearly not power laws and we tried to fit them with
log-normal functions. Values of coefficients are reported in table
\ref{tab:reg_cdf}. The curve for the in-degree is systematically above the one for
the out-degree. Given the fact that the pdf is the derivative of the cdf, the plot
implies that in the range [1 50] the in-degree is typically smaller than the
out-degree while above 50 the opposite holds. We don't have an explanation for
this finding. On the other hand we remind that as in the case of firms, the number
of connections is not necessarily meaningful. In fact this is exactly why we have
introduced the inward and outward investment stock.

\subsection{Regions. Activity, investment and connectivity degree: correlations}\label{sec:Results:corr_reg}
\label{sec:Results:correl_reg}

\begin{figure}[h]
    \begin{center}
      \includegraphics[width=0.47\textwidth]{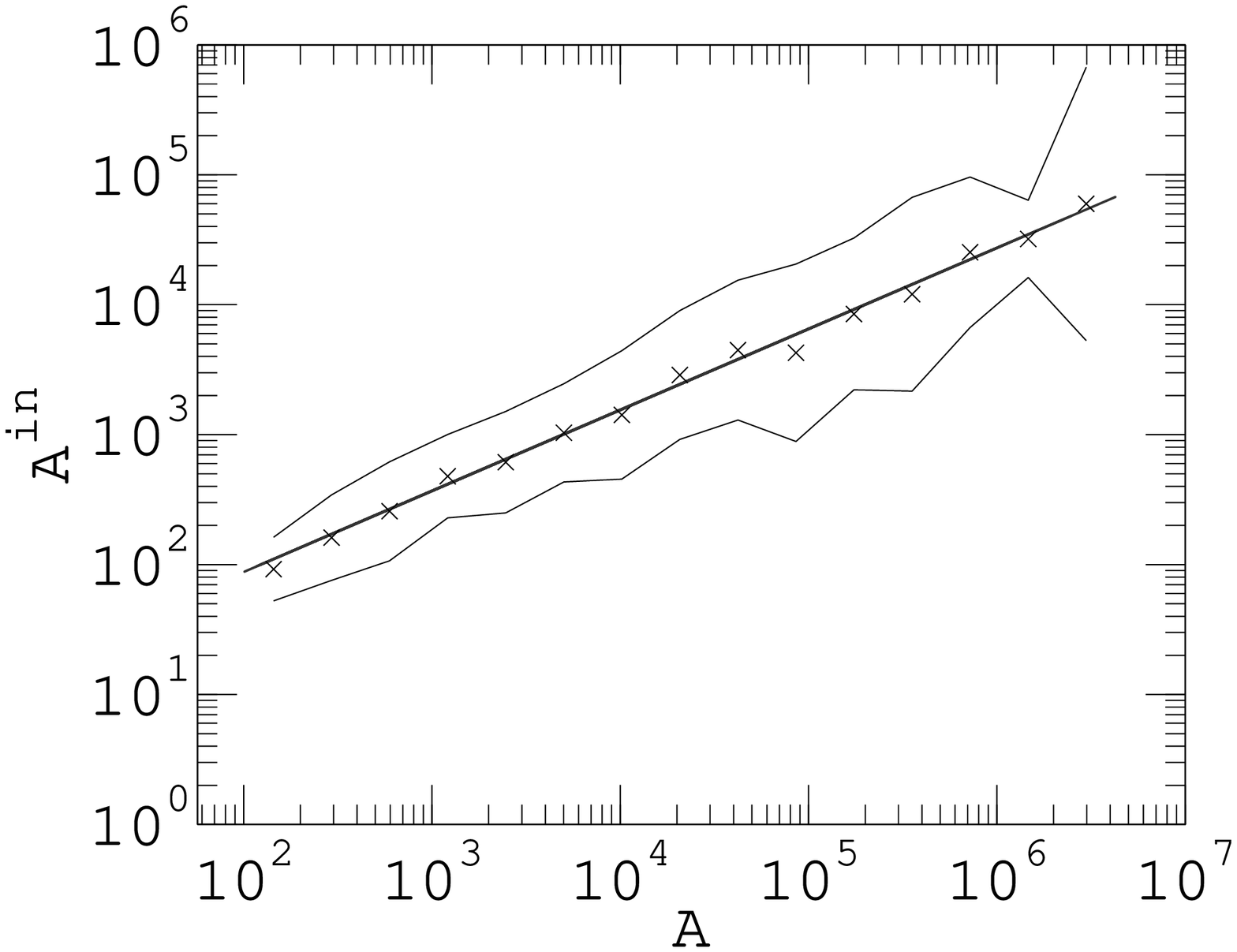}
      \hfill
      \includegraphics[width=0.47\textwidth]{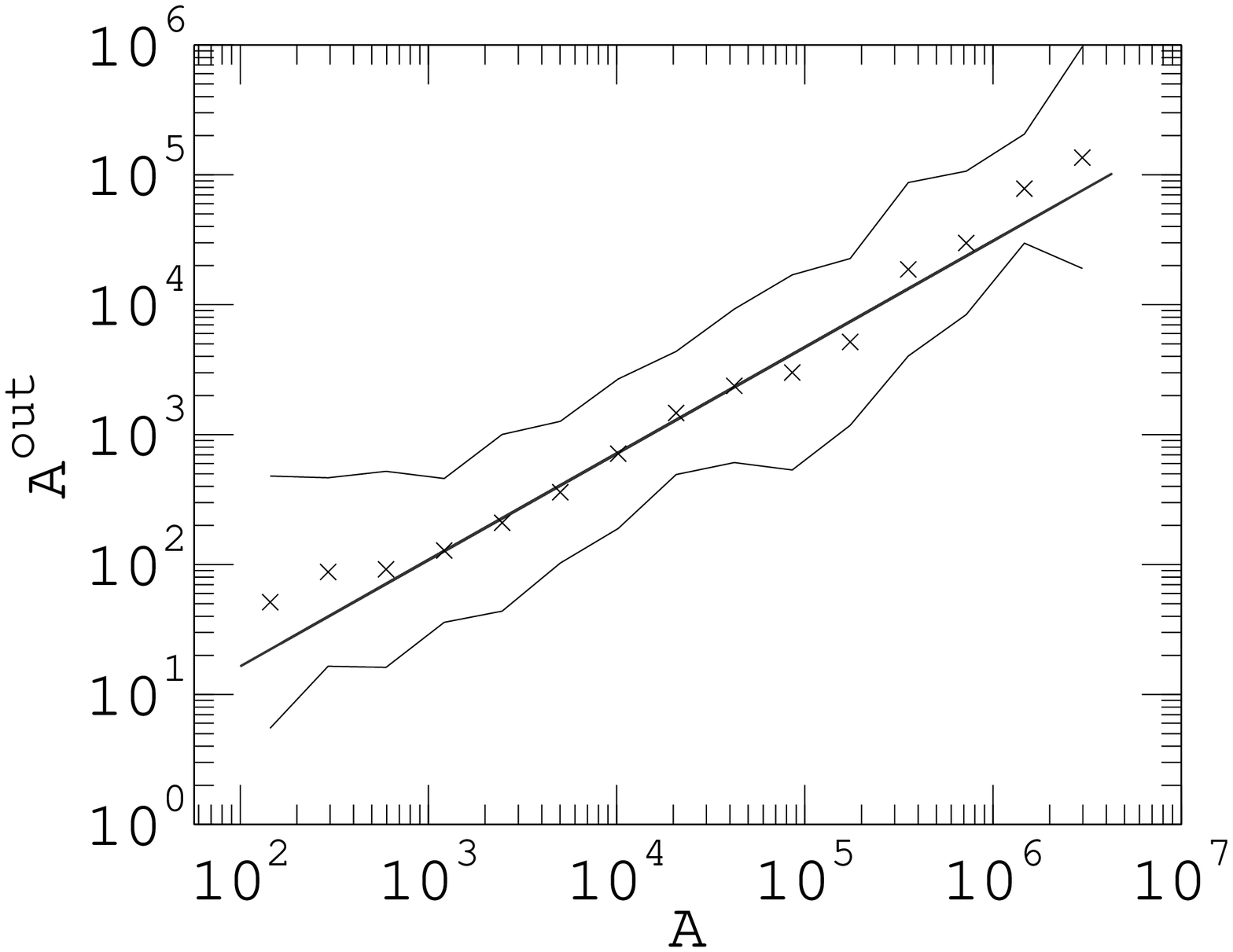}
      \caption{Regions. Plot of (left) hosted invested stock $A^{in}$ and (right)
      outward invested stock $A^{out}$ of regions versus their activity $A$ in
      $\log-\log$ scale. Data are binned.
      Mean $\pm$ standard deviation of the values in each bin are plotted as the continuous lines.
      \label{fig_reg_corr_in_out}}
    \end{center}
\end{figure}

\begin{figure}[h]
    \begin{center}
      \includegraphics[width=0.47\textwidth]{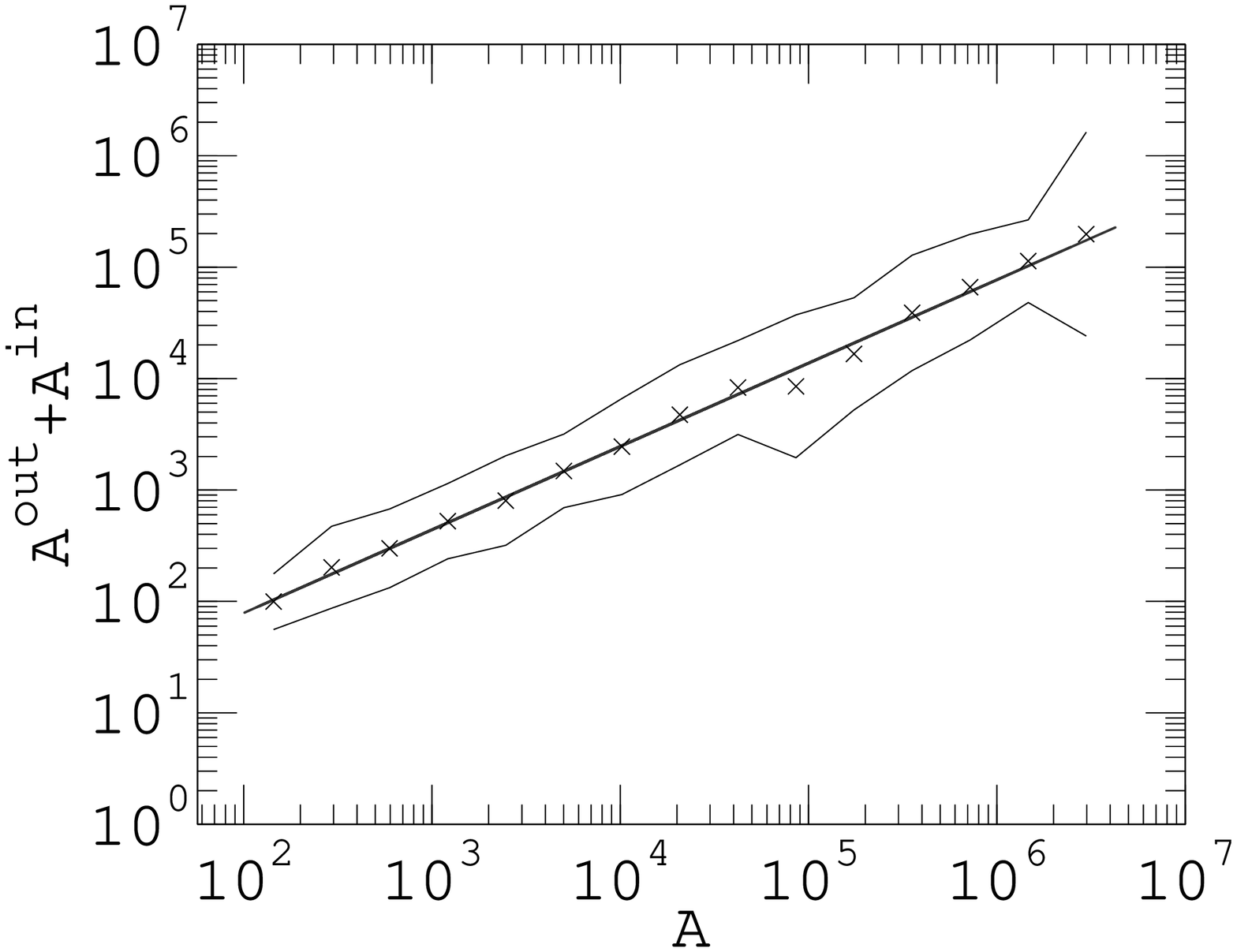}
      \hfill
      \includegraphics[width=0.47\textwidth]{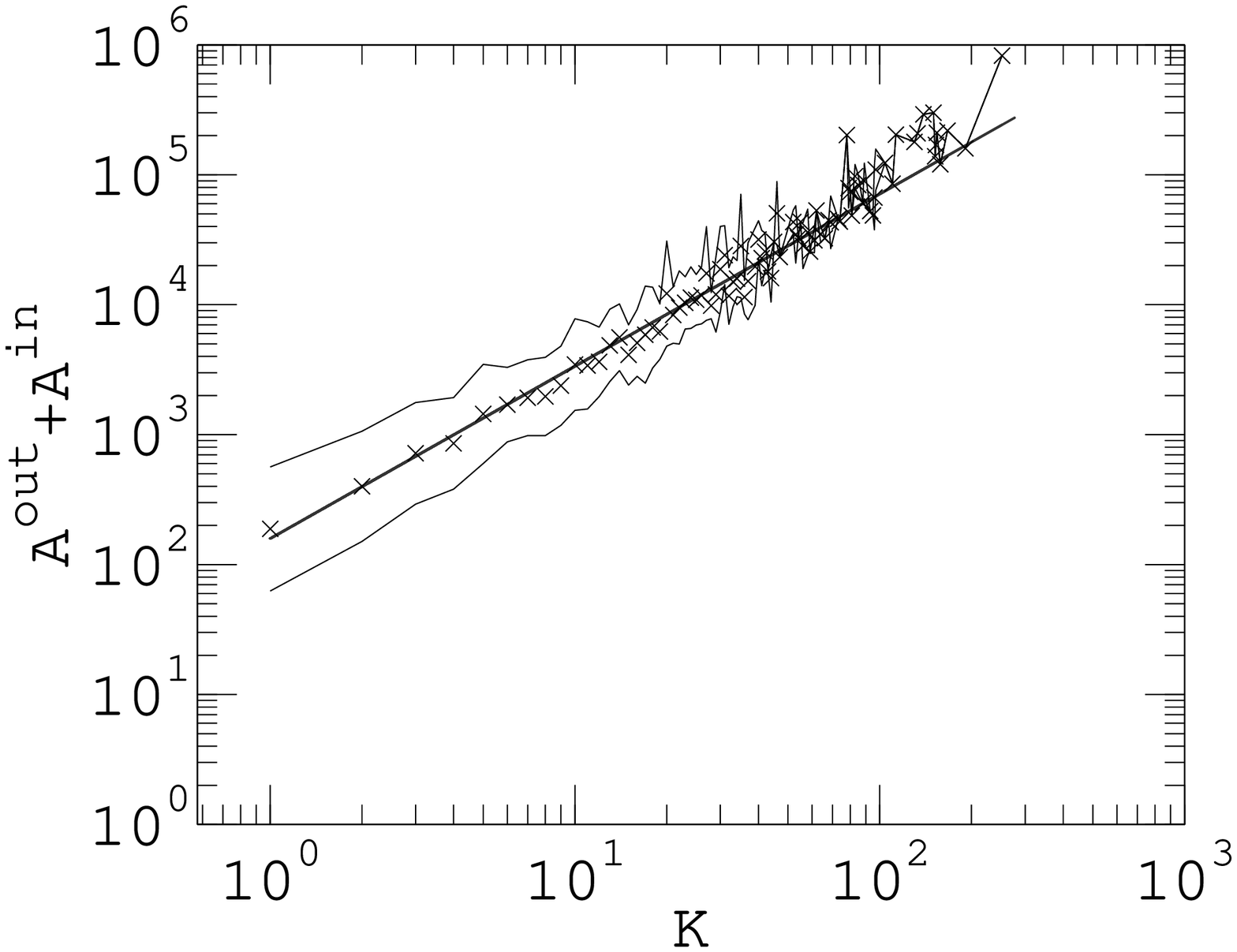}
      \caption{Regions. Plot of (left) total invested stock $A^{in+out}$ of
      regions versus their activity $A$. (right) Plot of total connectivity degree $k^{tot}$ of
      regions versus their hosted investment stock $A^{in+out}$. $\log-\log$ scale. Data are binned.
      Mean $\pm$ standard deviation of the values in each bin are plotted as the continuous lines.
      \label{fig_reg_corr_inout_ktot}}
    \end{center}
\end{figure}

In order to understand why investment of regions are log normal distributed and
why distributions of inward/outward investment and activity display a partial
overlap after normalization (see section \ref{sec:Methods:remarks}), we
investigate the correlations between activity, investment and connectivity
in/out-degree of regions. With the same procedure used for the case of firms
(see section \ref{sec:Results:correl_comp}), we produced the plots in figures
\ref{fig_reg_corr_in_out}-\ref{fig_reg_corr_inout_ktot} (see section
\ref{sec:Methods:remarks}.

\begin{figure}[h]
    \begin{center}
      \includegraphics[width=0.47\textwidth]{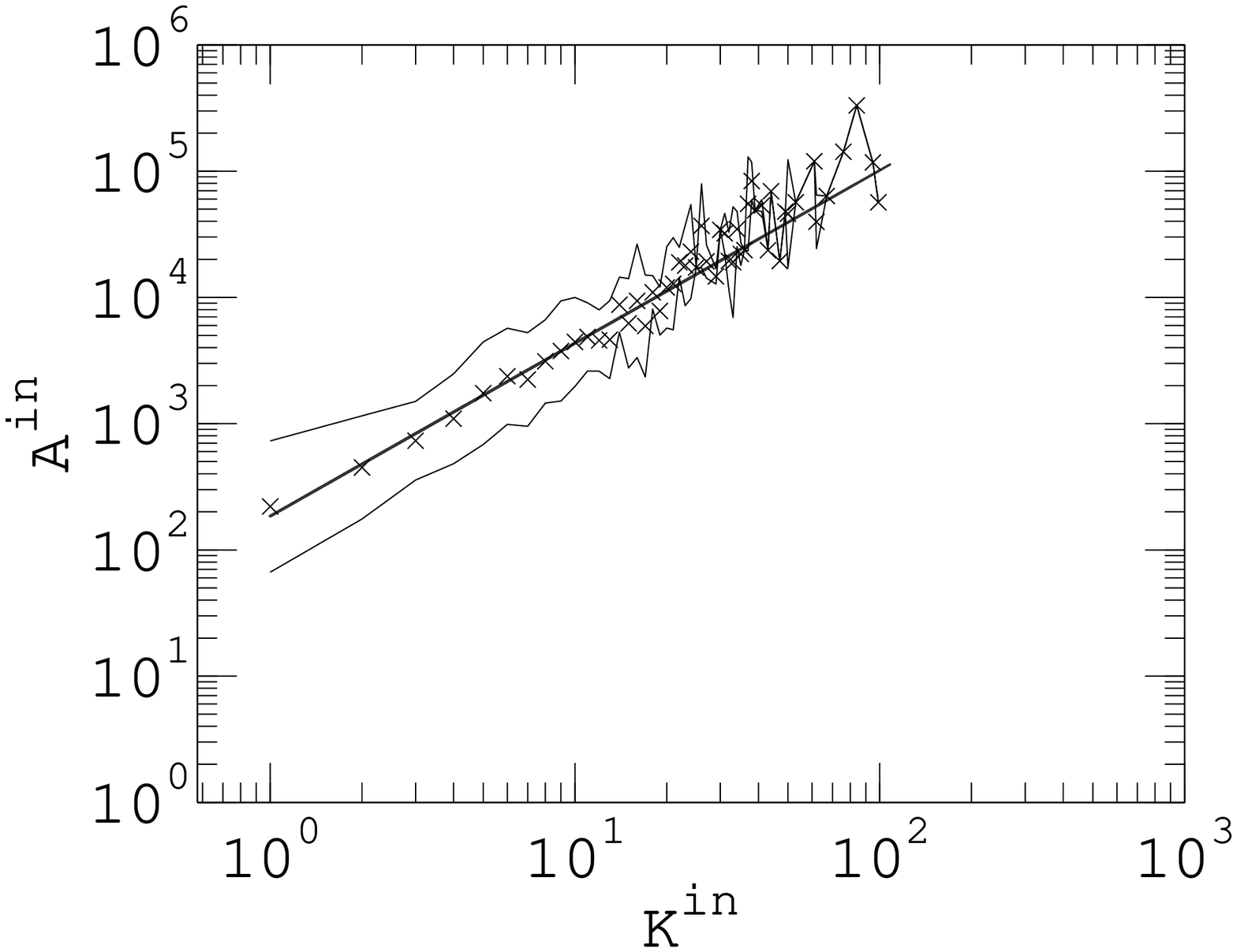}
      \hfill
      \includegraphics[width=0.47\textwidth]{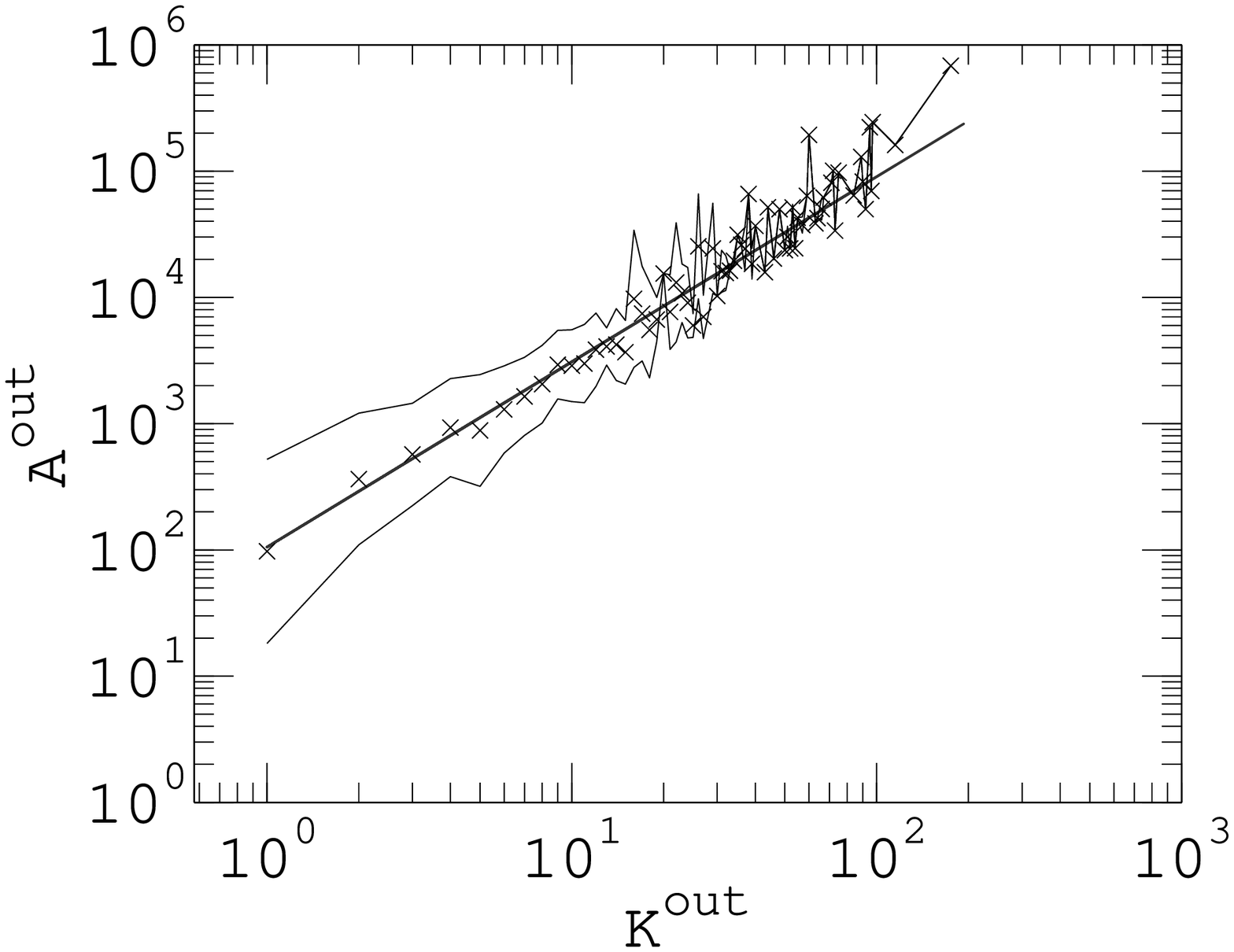}
      \caption{Regions. Plot of connectivity (left) in-degree $k^{in}$ and (right) out-degree of
      regions versus their hosted investment stock $A^{in}$ and outward investment stock $A^{out}$
      respectively in $\log-\log$ scale. Data are binned.
      Mean $\pm$ standard deviation of the values in each bin are plotted as the continuous lines.
      \label{fig_reg_corr_kin_kout}}
    \end{center}
\end{figure}

Tables \ref{tab_reg_corr},\ref{tab:tab_reg_kcorr} report the values of slope and
correlation coefficients for the linear fit of the binned data. Again the
correlation coefficients are all quite close to 1 so, with the caveat mentioned in
section \ref{sec:Methods:remarks}, we conclude that the data indicate the
existence of scaling laws in the network of regions between investment and
activity and between connectivity degree and activity.

\begin{table}[h]
\caption{Scaling of investments versus activity for regions.
 Coefficients of linear fit and correlation coefficients.
\label{tab_reg_corr}} {\begin{tabular}{@{}ccccc@{}} \toprule
  Yaxis      &   $m$      &     $q$   &    corr.coef.    \\ \colrule
$A^{in}$     &   0.623   &  0.697   &    0.997        \\
$A^{out}$    &   0.819   & -0.420   &    0.990        \\
$A^{in+out}$ &   0.748   &  0.401   &    0.998        \\ \botrule
\end{tabular}}
\end{table}

\vspace{1cm}
\begin{table}[h]
\caption{Scaling of investments versus connectivity degree for regions.
 Coefficients of linear fit and correlation coefficients.
\label{tab:tab_reg_kcorr}}
{\begin{tabular}{@{}ccccc@{}} \toprule
  Yaxis       &   $m$      &     $q$   &    corr.coef.    \\ \colrule
 $A^{in}$     &   1.467   &  2.022   &    0.965        \\
 $A^{out}$    &   1.370   &  2.268   &    0.965        \\
 $A^{in+out}$ &   1.326   &  2.201   &    0.979        \\ \botrule
\end{tabular}}
\end{table}

We notice that the exponents $0.62$ and $0.82$ for the scaling of hosted
investment versus activity and outward investment versus activity respectively are
smaller than 1.

As seen at the firm level, it follows that although more active regions tend
to make and host more investment, the investment increases less than linearly as a function of the
activity (sub-linear increase). This means that very active regions invest, in proportion,
less than smaller regions. However, the
increase of outward investment with activity is stronger for regions ($m=0.82$)
than for firms ($m=0.61$ see table \ref{tab:firm_corr}). This is an
interesting result that we will address in the future.

The amount of investment made or received by regions in relation with their
activity is relevant to institutions in charge of fostering development of
regions. Although this findings cannot provide detailed prediction they could help
develop multi-agent based models trying to reproduced the observed features with
the aim of designing possible incentive strategies.

Finally, the exponents for the scaling of activity versus in/out-degree are again
not far from the values found for the scaling law in other works such as: between invested volume and
degree in stock markets and for the scaling law of node strength versus degree in
air traffic networks.

\section{Conclusions}\label{sec:Conclusions}

In this paper we propose a simple but novel procedure to build the network of
inter-regional investment, based on the number of employees of firms in each region
and their network of ownership. In this network representation, the connectivity
in-degree of a region is the number of other regions from which firms invest in
the focal region, while the out-degree is the number of regions in which firms of
the focal region invest. The sum of the weights over the incoming links represents
the hosted investment stock of a region in terms of employees. The sum over the
outgoing links represents the outward investment stock of a region in terms of
employees.

We study the statistical properties of investment stock networks at the level of
firms and at the level of regions. Our first result is that investment stock of
firms is power law distributed and that, in particular, the exponent of outward
investment is very close to the one of firm activity. As it is well known, this
fact may result from a power law scaling relation between activity and outward
investment. This is neither obvious nor documented in the literature, so it has to
be checked empirically. At a first sight, activity and investment are quite
scattered and span a few orders of magnitude. However, by taking the logarithm of
the values of activity and investments, and binning the data, we indeed find that
investment scales as a power of the activity. Moreover, power law scaling
relations also hold between investment stock and connectivity degree.

On the other hand, in the case of regions, we find log normal distributions for
activity, investments and degree. Now, it can be argued that this result might
simply be related to the distribution of population size across regions
(unfortunately we do not have data to test this hypothesis at the moment). Even
so, it is a remarkable fact that such probability distributions for the regions
clearly differ both from those of firms as from those of countries in the world.
The impact of this fact on the design of global policies to foster investments and
economic development should be investigated.

Again, the fact that similar distributions emerge for activity, investment and
connectivity degree of regions suggests that some relation should hold among them.
In particular, we find that investment of firms scales as a power of the degree
with exponent 1.57 (out-degree), while for regions it scales with exponents 1.37
and 1.47 (in-degree and out degree respectively). Interestingly, previous studies
on different data set have investigated the scaling law of two quantities
(invested volume and node strength) strictly analogous to the investment stock and
found exponent values between 1.1 and 1.7
\cite{Barrat/Barth/Vesp:05:SpatConstrInAirNet,Garlaschelli/Battiston/al:05:SFTopo}.
The existence of scaling laws relating investment, activity and connectivity both
in firms and regions is an interesting and novel result relevant to the fields of
complex networks, industrial economics and geography.

On the other hand, such scaling laws should be of interest for policy making. For
instance we find that very active regions invest less, in proportion, with respect
to smaller ones. The same holds for firms, although the coefficient governing the
relation between investment and activity are different. This kind of result allows
to make some statistical predictions about the investments that regions will
receive or make, based on their activity and connectivity. The present work is a
first step towards understanding the relation between the local dynamics of
investment flows and the macro-economical facts emerging at a global level. One
can ask for example whether such a distribution of investments is desirable with
respect to some societal goals that might be at stake at the country or at the EU
level. If it is not, one can investigate if introducing some incentive policies,
the distribution can be improved with respect to the goals. In this sense, the
findings reported here should stimulate the investigation of models for managing
the development of regions and the optimal allocation of resources. Overall, we
believe that these results open the way for further studies with potential long
run implications in policy making at the level of EU investment promotion, support
to underdeveloped EU regions and optimization of investment flow.


\section{Acknowledgments}\label{sec:Acknowledgments}
This manuscript greatly benefited from Prof. Frank Schweitzer's comments. We also
thank Prof. Domenico Delli Gatti of Universit\`{a}  Cattolica di Milano for
granting us access to Amadeus database. S.B. acknowledges the support of CNRS
during the phase of data collection and the initial stage of this work. S.B. and
J.R. acknowledge the support of the European project MMCOMNET (IST contract n.
12999) and the EXYSTENCE Thematic Institute CHIEF (held at Univ. Politecnica delle
Marche, Ancona, IT, May 2-21 2005) where this collaboration has started.


\begin{thebibliography}{99}

\vspace*{-5pt}   

\bibitem{Axtell:01:Science} Axtell, R.L., Zipf Distribution of U.S. Firm Sizes, Science 293 (2001) 1818.

\bibitem{Barrat/Barth/Vesp:05:SpatConstrInAirNet}
Barrat, A., Barth\'{e}lemy, M., Vespignani, A., The effects of spatial constraints
on the evolution of weighted complex networks", arXiv:physics/0504029 v1 4 Apr
2005

\bibitem{Basile/Castellani/Zanfei:03:LocationChoiceMultinat}
Basile, R., Castellani, D., Zanfei, A., Location choices of multinational firms in
Europe: the role of national boundaries and EU policy, Quaderni di economia,
matematica e statistica , n. 76, June 2003, (http://www.fscpo.unict.it/pdf/Castellani

\bibitem{Battiston:04:InnerStructCapitNets}
Battiston, S., Inner structure of capital control networks, Volume 338, Issues
1-2, 1 July 2004, Pages 107-112

\bibitem{Borensztein/DeGregorio/Lee:98}
Borensztein, E., De Gregorio, J., Lee, J.W., "How does foreign direct investment
affect economic growth?, Journal of International Economics 45 (1998) 115-135

\bibitem{Fujiwara/DiGuilmi/Aoyama/Gallegati:03:ParetoZipf}
Fujiwara, Y., Di Guilmi, C., Aoyama, H., Gallegati, M., Souma, W., Do Pareto-Zipf
and Gibrat laws hold true? An analysis with European Firms, arXiv:cond-mat/0310061
v2 27 Nov 2003

\bibitem{Garlaschelli/Loffredo:04:WTW}
Garlaschelli, D., Loffredo, M.I., Fitness-dependent topological properties of the
World Trade Web, Phys. Rev. Lett. 93, 188701 (2004), arXiv:cond-mat/0403051

\bibitem{Garlaschelli/Battiston/al:05:SFTopo}
Garlaschelli, D., Battiston, S., Castri, M., Servedio, V.D.P., Caldarelli, G., The
scale free topology of market investments, 2005, Physica A, Volume 350, 2-4,
491-499

\bibitem{Gaffeo/Gallegati/Palestrini:03}
Gaffeo, E., Gallegati, M., Palestrini, A., Physica A 324 (2003) 117.

\bibitem{DelliGatti/DiGuilmi/Gallegati:03}
Delli Gatti, D., Di Guilmi, C. and Gallegati, M., On the Empirics of "Bad Debt",
mimeo, Università Politecnica delle Marche (2003).

\bibitem{Delli Gatti/al:03:NewApprBusinessFluct}
Delli Gatti, D., Di Guilmi, C., Gaffeo, E., Giulioni, G., Gallegati, M.,
Palestrini, A., A new approach to business fluctuations: heterogeneous interacting
agents, scaling laws and financial fragility, to appear on Journal of Economic
Behaviour and Organisation, http://arxiv.org/abs/cond-mat/0312096

\bibitem{Goldstein/Morris/Yen:04}
Goldstein, M.L., Morris, S.A., and Yen G.G., Problems with Fitting to the
Power-Law Distribution, to appear on EPJ, arXiv:cond-mat/0402322 v3 13 Aug 2004


\bibitem{Guimera/Mossa/Turtschi/Amaral:05:AirNet}
Guimer\`{a} R., Mossa S., Turtschi A., Amaral L.A.N., The world-wide air
transportation network: Anomalous centrality, community structure, and cities'
global roles", Proc. Nat. Acad. Sci. USA 102, 7794-7799 (2005)

\bibitem{Kesten:73:RandomDiff} Kesten, H., Random di®erence equations and renewal theory for products of
random matrixes. Acta Math., 131, 207-248,(1973)

\bibitem{Kogut and Walker 1999}
Kogut, B. and Walker, G., The Small World of Germany and the Durability of
National Ownership Networks, American Sociological Review, 66(3): 317-35, 2001.

\bibitem{Jakobsen/Onsager:04:HeadOfficeLocation}
Jakobsen, S.E. and Onsager, K., Head office location – Agglomeration, clusters or
flow nodes?  Urban Studies,  August, 2005 Volume 42:9

\bibitem{Loungani/Razin}
Loungani, P. and Razin, A., "How Beneficial Is Foreign Direct Investment for
Developing Countries?" Finance and Development (IMF) June 2001, Volume 38, Number
2


\bibitem{Newman:05:PLReview}
Newman, M. E. J., Power laws, Pareto distributions and Zipf's law,
arXiv:cond-mat/0412004 v2 9 Jan 2005


\bibitem{Serrano/Boguna:03:WTW}
Serrano, M.A. and Bogu\~{n}\'{a}, M., Topology of the World Trade Web, Physical
Review E 68, 015101 (2003), arXiv:cond-mat/0301015

\bibitem{Sornette:98}
Sornette, D., Multiplicative processes and power laws, Phys. Rev. E 57 N4,
4811-4813 (1998)

\bibitem{Sutton:97}
Sutton, J., J. Econ. Lit. 35 (1997) 40.

\bibitem{Wallerstein:74:ModernWorldSystemI}
Wallerstein, I., The Modern World System I. Capitalist Agriculture and the Origins
of the European World -Economy in the Sixteenth Century, New York, Academic Press,
1974

\bibitem{ONS-UNCTAD:FDI_def}
Office for National Statistics (ONS), UK - FDI Profile UNCTAD

\bibitem{IFC:Policies}
Policies to Attract Foreign Direct Investment, World Bank/IFC. (Web site: Papers
and Links - Privatization, Infrastructure \& Business Environment)

\end{thebibliography}
\end{document}